\documentclass[epjST]{svjour}
\usepackage{graphics}
\usepackage{tabularx}
\usepackage{xcolor}
\usepackage[nosort]{cite}
\usepackage[T1]{fontenc}
\usepackage{epsfig}
\usepackage{latexsym}
\usepackage{graphicx}
\usepackage{amsmath}
\usepackage{amsfonts}   
\usepackage{amssymb}    
\usepackage{float}
\usepackage{bm}
\usepackage{url}
\usepackage{hyperref} 
\usepackage[nodisplayskipstretch]{setspace}
\def\lsim{\raise0.3ex\hbox{$\;<$\kern-0.75em\raise-1.1ex\hbox{$\sim\;$}}}
\def\gsim{\raise0.3ex\hbox{$\;>$\kern-0.75em\raise-1.1ex\hbox{$\sim\;$}}}

\newcommand{\captions}{\sf\caption}
\def    \beq            {\begin{equation}}
\def    \eeq            {\end{equation}}
\def    \bea           {\begin{eqnarray}}
\def    \eea           {\end{eqnarray}}

\def \mn{\mu\nu{\rm SSM}}

\def\g2{{\rm GeV}^2}
\def\tb{{\rm tan}\beta}

\def\sw2{sin^2 \theta_w}

\def\a^tau{\alpha_{\tau}}

\def\beq{\begin{equation}}
\def\eeq{\end{equation}}
\def\beqa{\begin{eqnarray}}
\def\eeqa{\end{eqnarray}}

\newcommand{\newc}{\newcommand}
\newc\BR{BR}
\newc{\akappa}{A_{\kappa} }
\newc\deltagmtwo{\delta (g-2)_{\mu}} 
\newc\deltaamu{\Delta a_{\mu}}

\def\anti{\overline}

\def\la{\lambda}
\def\bla{\bm \lambda}

\newc{\haa}{BR\(h_1\to a_1 a_1\)}
\newc{\abb}{BR\(a_1\to b\anti{b}\)}
\newc{\hbb}{BR\(h_1\to b\anti{b}\)}
\newc{\abund}{\Omega h^2}
\newc\bsgamma{b\rightarrow s \gamma }
\newc\bxsgamma{\overline{B}\rightarrow X_{s}\gamma}
\newc\brbsgamma{\BR(\overline{B}\rightarrow X_s\gamma)}

\newc{\Fermi}{\textit{Fermi}-}

\allowdisplaybreaks

\begin{document}
\title{Searching for Supersymmetry: The $\mu\nu$SSM}
\subtitle{A short review}
\author{Daniel~E.~L\'opez-Fogliani\inst{1}\fnmsep\inst{2}\fnmsep\thanks{\email{daniel.lopez@df.uba.ar}} \and Carlos Mu\~noz\inst{3}\fnmsep\inst{4}\fnmsep\thanks{\email{c.munoz@uam.es}}}
\institute{
Instituto de F\'isica de Buenos Aires UBA \& CONICET, Departamento de F\'isica,
 Facultad de Ciencia Exactas y Naturales, 
Universidad de Buenos Aires, 
1428 Buenos Aires, Argentina
\and Pontificia Universidad Cat\'olica Argentina, 
1107 Buenos Aires, Argentina
 \and 
Departamento de F\'isica Te\'orica, Universidad Aut\'onoma de Madrid (UAM),
Campus de Cantoblanco, 28049 Madrid, Spain
\and
Instituto de F\'{\i}sica Te\'{o}rica UAM-CSIC, 
  Campus de Cantoblanco, 28049 Madrid, Spain}
\abstract{We review the role played by the `$\mu$ from $\nu$' supersymmetric standard 
model ($\mu\nu$SSM) in the search for supersymmetry. First, we discuss its theoretical motivation, that is the simultaneous solution of $\mu$- and $\nu$-problems through the introduction of right-handed neutrinos. The latter produces $R$-parity violation (RPV), giving rise to interesting
signals of new physics. As by-products, in the $\mu\nu$SSM there are dark matter candidates, 
and electroweak baryogenesis can be realized. Then, we survey signals by which the model could be tested at the large hadron collider (LHC). In addition to the enlarged Higgs sector with sneutrinos, we put special emphasis in analyzing 
the {intimate} connection between the lightest supersymmetric particle (LSP) lifetime and the size of neutrino Yukawa couplings. Displaced vertices and/or multileptons are some of the interesting signatures that can be probed. Finally, we discuss possible extensions of the $\mu\nu$SSM such as the inclusion in the superpotential of {the} conventional trilinear lepton-number violating couplings, the addition of an extra $U(1)'$ gauge group to the symmetry of the standard model, or 
the reinterpretation of the Higgs doublets as a fourth family of leptons superfields
motivating the existence of a fourth family of vector-like quark doublet superfields.
} 
\maketitle     %

\section{Introduction}%
\label{intro}         %
Supersymmetry (SUSY)~\cite{Golfand:1971iw,Volkov:1972jx,Wess:1974jb} is one of the most compelling frameworks for physics beyond the standard model (SM), providing also an elegant solution to the gauge hierarchy problem. The minimal SUSY extension of the SM is known as the minimal supersymmetric standard model (MSSM)(for reviews, see e.g. Refs.~\cite{Nilles:1983ge,Haber:1984rc,Martin:1997ns}). 
However, neutrinos are massless in this model implying that the MSSM itself is unable to solve the 
$\nu$-problem, i.e. the generation of neutrino masses according to experimental results~\cite{Capozzi:2017ipn,deSalas:2017kay,deSalas:2018bym,Esteban:2018azc,Abe:2019vii}. The MSSM suffers also a naturalness problem, the so-called 
$\mu$-problem~\cite{Kim:1983dt}  (for a recent review, see Ref.~\cite{Bae:2019dgg}). This arises from the requirement of a mass term for the Higgs superfields in the 
superpotential, $\mu \hat H_d \hat H_u$, which must be of the order of the electroweak (EW) scale to successfully lead to electroweak symmetry breaking (EWSB), as well as to generate Higgsino masses compatible with current experimental lower bounds~\cite{Tanabashi:2018oca} on SUSY particles (sparticles)~\cite{Fayet:1976et,Fayet:1977yc}.
However, the natural scale for $\mu$, being a SUSY mass term, is the high-energy scale expected in the theory which can be $M_{\text{GUT}}$, $M_{\text{string}}$, {or,
in the absence of a unified theory, $M_{\text{Planck}}$, since gravity is always there}. In the MSSM, there is no attempt to solve this problem, the $\mu$ term is just assumed to be present. 

The `$\mu$ from $\nu$' supersymmetric standard 
model ($\mu\nu$SSM)~\cite{LopezFogliani:2005yw} is a natural alternative to the MSSM 
which {\bf solves dynamically both the $\mu$- and $\nu$-problems}.
\footnote{More details about the model can be found in the website of the $\mn$ Working Group, http://dark.ft.uam.es/mununiverse}
In this framework,
the only three possible types of gauge-invariant trilinear couplings involving right-handed (RH) neutrinos $\nu_{iR}$
are added to the superpotential $W$ containing the usual Dirac Yukawa couplings for quarks and charged leptons.
First, one adds 
couplings between 
$\nu_{iR}$ 
and Higgses of the type $\lambda_{i} \, \hat \nu^c_i\,\hat H_d \hat H_u$,
where $\hat \nu^c_i$
denote the RH neutrino superfields. 
After the successful EWSB, 
the right sneutrinos 
$\widetilde \nu_{iR}$ develop vacuum expectation values (VEVs) of the order of TeV, and therefore 
the above couplings generate an effective {\bf mass term for Higgsinos},
$\mu
=\la_i \langle \widetilde \nu_{iR}\rangle^*$,
solving the $\mu$-problem. Note in this regard that, although all couplings in $W$ are dimensionless, the EWSB is
driven by the soft SUSY-breaking terms which are in the ballpark of a TeV.
In addition,
Dirac Yukawa couplings for neutrinos $Y^{\nu}_{ij} \, \hat H_u\, \hat L_i \, \hat \nu^c_j$ are also added, as well as
couplings between $\nu_{iR}$ themselves
$\kappa{_{ijk}} \hat \nu^c_i\hat \nu^c_j\hat \nu^c_k$.
The latter couplings,
in addition to forbid a global U(1) symmetry in $W$ avoiding the existence of a Goldstone boson, 
generate effective {\bf Majorana masses for RH neutrinos},
$
{\mathcal M}
_{ij}={2}\kappa_{ijk}\langle \widetilde \nu_{kR}\rangle^*$. Both types of couplings
are therefore 
instrumental 
in solving the $\nu$-problem through a {\bf EW-scale seesaw}: they
can accommodate with
$Y^{\nu}_{ij} \lsim 10^{-6}$~\cite{LopezFogliani:2005yw,Escudero:2008jg,Ghosh:2008yh,Bartl:2009an,Fidalgo:2009dm,Ghosh:2010zi,Liebler:2011tp} the correct neutrino masses and mixing angles. 
Having 
a {EW seesaw} also avoids the introduction of {\it ad-hoc} high-energy scales in the model, as it occurs e.g. in the case of a GUT seesaw.  
Thus, the only scale in the $\mn$ is the EWSB scale.
It is worth noticing here that the $\mn$ seesaw is actually a {\it generalized} EW seesaw, because it involves not only left-handed (LH) neutrinos $\nu_{iL}$ with RH ones 
$\nu_{iR}$ as in the usual seesaw, but also the neutralinos.
This fact favors the accommodation of neutrino data, and occurs because of the RPV
present in the $\mn$, which we are going to discuss now.


The $Z_2$ discrete symmetry $R$-parity 
($+1$ for particles and $-1$ for sparticles)~\cite{Farrar:1978xj} is used nowadays in SUSY in order to avoid fast proton decay induced by the simultaneous presence of the lepton- and baryon-number violating couplings~\cite{Weinberg:1981wj,Sakai:1981pk} 
$\lambda'_{ijk}\hat L_i \hat Q_{j} \hat d^{c}_{k}$ 
and $\lambda''_{ijk}\hat d^c_{i} \hat d^c_{j} \hat u^c_{k}$, 
respectively.
Imposing $R$-parity conservation (RPC), those couplings are forbidden.
Also, the LSP turns out to be stable since RPC establishes that sparticles appear in the couplings always in pairs. Being stable, as a by-product the LSP contributes to dark matter (DM). Therefore, it has to be neutral, as it is the case of the neutralino in the MSSM. 
As another consequence, the main signal of the RPC MSSM at colliders is missing transverse energy (MET).
Unlike the previous situation, in the $\mn$ the simultaneous presence of the above three types of couplings involving 
RH neutrinos, $\lambda_i, \kappa_{ijk}$ and $Y^{\nu}_{ij}$, produces 
{\bf explicit RPV and lepton-number violation}. 
Despite these effects, it is important to note the following.
First of all, these couplings 
are harmless for proton decay. 
Second, lepton-flavor violation is very small and therefore harmless in the $\mn$, fulfilling this kind of processes all
the experimental constraints.
The reason is simple.  
In the limit $Y^\nu_{ij} \to 0$
only $\lambda_i$ and $\kappa_{ijk}$ couplings are left, and therefore
$\hat \nu^c_i$ can be identified as
pure singlet superfields without lepton number. 
Thus $Y^\nu_{ij}$ are the parameters determining lepton-number violation,
and such violation is small since
$Y^{\nu}_{ij} \lsim 10^{-6}$.
For the same reason, RPV is also small in the $\mn$. {However, although small, such violation} has crucial implications concerning the phenomenology. 
RPV implies that the
LSP\footnote{The notion of LSP is in fact misleading in the context of RPV models, since SUSY and non-SUSY states are mixed. Nevertheless, for dominant SUSY composition of the lightest eigenstate, it is reasonable to keep this nomenclature, as we will do in what follows.} 
is not stable, decaying into SM particles. 
As a consequence, in the $\mn$ {\bf the smallness of neutrino masses is directly related with the low decay width of the LSP}.
Actually, it is also related {to} the existence of {\bf SUSY decaying DM candidates} in the model.
This is the case of 
the gravitino~\cite{Choi:2009ng,GomezVargas:2011ph,Albert:2014hwa,GomezVargas:2017,Gomez-Vargas:2019mqk}, or the axino~\cite{Gomez-Vargas:2019vci}, with {lifetimes} greater than the age of the Universe.
The key to this is that gravitino (axino) interactions are suppressed not only by the Planck (Peccei-Quinn) scale, but also by the small RPV parameters. 
It is also worth mentioning {concerning} cosmology, that baryon asymmetry might be realized in the
$\mn$ through {\bf electroweak baryogenesis}~\cite{Chung:2010cd}.

The search for low-energy SUSY is one of the main goals of the LHC. This search
has been focused mainly on prompt signals with MET inspired in RPC models, such as the MSSM. 
In that framework, significant bounds on sparticle masses have been obtained~\cite{Tanabashi:2018oca}. For strongly interacting sparticles, their masses must be above about 1 TeV. For weakly interacting sparticles, the lower bounds are of about 100 GeV, and only the bino-like neutralino is basically not constrained due to its small pair production cross section.
Because of these results, there is a growing interest in searching for {\bf displaced signals at the LHC} (for recent review, see Ref.~\cite{Alimena:2019zri}). The predictions of the $\mn$ are in this sense robust, and point to a coherent search strategy. 
All sparticles are basically potential candidates for the {LSP}, not only the neutral ones as in the RPC MSSM.
This means that in the $\mn$,  
{\bf squarks, gluinos, charginos, charged sleptons, sneutrinos,
or neutralinos can be the LSP, 
decaying prompt or with a decay length of the order of mm$-$m}, depending on the sparticle couplings, {their masses}, and the region of the parameter space analyzed.
These features have two crucial consequences. First, they imply that the extrapolation of the usual bounds on sparticle masses to the $\mn$ is not applicable. 
Second, the $\mn$
can produce distinctive signals
of the model at colliders involving 
light sparticles which decay producing multi-leptons/jets/photons with small/moderate MET from neutrinos~\cite{Ghosh:2008yh,Bartl:2009an,Bandyopadhyay:2010cu,Liebler:2011tp,Fidalgo:2011ky,Ghosh:2012pq,Ghosh:2014rha,Ghosh:2014ida,Ghosh:2017yeh,Biekotter:2017xmf,Lara:2018rwv,Lara:2018zvf,Biekotter:2019gtq,Kpatcha:2019gmq,Kpatcha:2019qsz}, 
well verifiable at the LHC or at {upcoming accelerator 
experiments.}
Also {verifiable} signals {might be produced} at non-collider experiments such as neutrino experiments or the 
muon $g-2$~\cite{Kpatcha:2019pve}. 

To carry out these analyses, it is crucial to take into account that 
because of RPV {\bf all fields in the spectrum of the $\mn$ with the same color, electric charge and spin mix together}.
The associated mass matrices were studied in detail in 
Refs.~\cite{Escudero:2008jg,Ghosh:2008yh,Bartl:2009an,Ghosh:2017yeh}.
For example, 
there is a substantial {\bf mixing between neutrinos and neutralinos}. This is the origin of the generalized EW seesaw of the $\mn$, discussed above. Also, there is a 
{\bf mixing between right sneutrinos and doublet-like Higgses}.
The searches for new Higgses and deviations from the SM predictions is an active topic at the LHC.
In the context of the $\mn$, 
it was found that large regions of the parameter space compatible with neutrino physics as well as with flavor observables such as $B$ and $\mu$ decays are viable, containing an interesting phenomenology that 
could be probed at the 
LHC~\cite{Escudero:2008jg,Bandyopadhyay:2010cu,Fidalgo:2011ky,Ghosh:2012pq,Ghosh:2014ida,Biekotter:2017xmf,Biekotter:2019gtq,Kpatcha:2019qsz}.
For example, {one might probe processes such as new two-body decays for the SM-like Higgs
into a pair of light scalars/pseudoscalars, neutralinos, or RH neutrinos, producing
final states with multi-leptons/jets/photons.}

Let us finally mention that there has been also studies of {\bf extensions of the $\mn$}, interesting both from the phenomenological and cosmological viewpoints.
First of all, one expects naturally the presence of {\bf conventional lepton-number violating couplings} $\lambda_{ijk} \hat L_i \hat L_j \hat e^c_k$ and $\lambda'_{ijk}\hat L_i \hat Q_{j} \hat d^{c}_{k}$ in the superpotential of the $\mn$, as was discussed in detail in 
Refs.~\cite{Escudero:2008jg,Lopez-Fogliani:2017qzj,Ghosh:2017yeh}.
Besides, the {\bf addition of an extra $U(1)'$} gauge symmetry in the $\mn$ was proposed 
in Ref.~\cite{Fidalgo:2011tm} to forbid the presence of dangerous operators in the superpotential, such as e.g. baryon-number violating couplings.
There, {its associated phenomenology containing a new $Z'$ gauge boson was also studied.}
Another extension of the $\mn$ was proposed in 
Ref.~\cite{Lozano:2018esg}, but allowing non-universal $U(1)'$ charges for the SM fields.
The $U(1)'$ extensions of the $\mn$ also motivate an extended SM singlet sector, where the number of singlets
is fixed by anomaly cancellation conditions~\cite{lopez:2020xx}. 
In this scenario, not only the phenomenology is modified, but
also the cosmology since {there are models where} some of the extra singlets can be
weakly interacting massive particles {\bf (WIMPs) contributing to stable DM}.
Finally, in 
Ref.~\cite{Lopez-Fogliani:2017qzj} {{\bf a reinterpretation of the Higgs field in SUSY} 
motivating the possibility of a fourth family of {\bf vector-like quark doublets}
was proposed, and their signals} were studied in 
Ref.~\cite{Aguilar-Saavedra:2017giu}.


The paper is organized as follows. In Sec.~\ref{the model}, we will briefly review the $\mn$, discussing its superpotential, soft terms, scalar potential and parameter space. On the way, we will describe the proton decay and naturalness problems which are common to all SUSY models, and our viewpoint to attack them. Finally, we will compare the main characteristics of the $\mn$ with those of other interesting models that are found in the literature.
In Sec.~\ref{sec:spectrum}, the spectrum of the model will be analyzed paying special attention
to the neutral fermion mass matrix which determines neutrino masses and mixing angles, and to the
neutral scalar sector where Higgses and sneutrinos are mixed.
{Collider and non-collider signals of the model that have been analyzed so far, will be discussed in Sec.~\ref{sec:searches}}.
{Whereas the EW} sector of the model has been studied in some detail, studies of the color sector are still missing. {We will see that given} the current analyses, the experimental constraints on the parameter space of the $\mn$ are still very mild.
In Sec.~\ref{sec:cosmo}, we will discuss the cosmology of the model. We will review the existence of SUSY candidates for decaying DM, such as the gravitino and/or the axino, and their possible detection in gamma-ray satellites.
Other DM candidates such as {sterile RH neutrinos} will be presented. We will also briefly discuss the generation of the baryon asymmetry through the mechanism of 
{electroweak baryogenesis}.
The potential cosmological domain wall problem is addressed at the end of the section.
In Sec.~\ref{sec:extensions}, we will briefly review the extensions of the $\mn$ proposed in the literature, and, in connection with this, a reinterpretation of the Higgs field in SUSY.
Our conclusions and prospects for future works are left for Sec.~\ref{conclusions}.
\section{The Model 
}
\label{the model}

As discussed in the Introduction, the simplest superpotential of the $\mn$~\cite{LopezFogliani:2005yw,Escudero:2008jg,Ghosh:2017yeh} with three RH neutrinos is the following:\footnote{For a detailed description of the superfield notation used, see Appendices A and B of Ref.~\cite{Ghosh:2017yeh}.} 
%
\bea
W = &&
\epsilon_{ab} \left(
Y^e_{ij} \, \hat H_d^a\, \hat L^b_i \, \hat e_j^c +
Y^d_{ij} \, \hat H_d^a\, \hat Q^{b}_{i} \, \hat d_{j}^{c} 
+
Y^u_{ij} \, \hat H_u^b\, \hat Q^{a}_{i} \, \hat u_{j}^{c}
\right)
\nonumber\\
&+&   
\epsilon_{ab} \left(
Y^{\nu}_{ij} \, \hat H_u^b\, \hat L^a_i \, \hat \nu^c_j 
-
\lambda_{i} \, \hat \nu^c_i\, \hat H_u^b \hat H_d^a
\right)
+
\frac{1}{3}
\kappa{_{ijk}} 
\hat \nu^c_i\hat \nu^c_j\hat \nu^c_k,
\label{superpotential}
\eea
where the summation convention is implied on repeated indexes throughout the paper unless explicitly specified, with $i=1,2,3$ the usual family indexes of the SM, and 
$a,b=1,2$ 
$SU(2)_L$ indexes.
$\epsilon_{ab}$ is the totally antisymmetric tensor $\epsilon_{12}=1$.

Working in the framework of a typical low-energy SUSY, the Lagrangian  containing the soft SUSY-breaking terms related to $W$ 
is given by:
\bea
-\mathcal{L}_{\text{soft}}  =&&
\epsilon_{ab} \left(
T^e_{ij} \, H_d^a  \, \widetilde L^b_{iL}  \, \widetilde e_{jR}^* +
T^d_{ij} \, H_d^a\,   \widetilde Q^b_{iL} \, \widetilde d_{jR}^{*} 
+
T^u_{ij} \,  H_u^b \widetilde Q^a_{iL} \widetilde u_{jR}^*
+ \text{h.c.}
\right)
\nonumber \\
&+&
\epsilon_{ab} \left(
T^{\nu}_{ij} \, H_u^b \, \widetilde L^a_{iL} \widetilde \nu_{jR}^*
- 
T^{\lambda}_{i} \, \widetilde \nu_{iR}^*
\, H_d^a  H_u^b
+ \frac{1}{3} T^{\kappa}_{ijk} \, \widetilde \nu_{iR}^*
\widetilde \nu_{jR}^*
\widetilde \nu_{kR}^*
\
+ \text{h.c.}\right)
\nonumber\\
&+&   
m_{\widetilde{Q}_{ij}}^2
\widetilde{Q}_{iL}^{a*}
\widetilde{Q}^a_{jL}
{+}
m_{\widetilde{u}_{ij}}^{2}
\widetilde{u}_{iR}^*
\widetilde u_{jR}
+ 
m_{\widetilde{d}_{ij}}^2
\widetilde{d}_{iR}^*
\widetilde d_{jR}
+
m_{\widetilde{L}_{ij}}^2
\widetilde{L}_{iL}^{a*}  
\widetilde{L}^a_{jL}
\nonumber\\
&+&
m_{\widetilde{\nu}_{ij}}^2
\widetilde{\nu}_{iR}^*
\widetilde\nu_{jR} 
+
m_{\widetilde{e}_{ij}}^2
\widetilde{e}_{iR}^*
\widetilde e_{jR}
+ 
m_{H_d}^2 {H^a_d}^*
H^a_d + m_{H_u}^2 {H^a_u}^*
H^a_u
\nonumber \\
&+&  \frac{1}{2}\, \left(M_3\, {\widetilde g}\, {\widetilde g}
+
M_2\, {\widetilde{W}}\, {\widetilde{W}}
+M_1\, {\widetilde B}^0 \, {\widetilde B}^0 + \text{h.c.} \right).
\label{2:Vsoft}
\eea
%
%
%
%
In case of following the usual assumption based on the breaking of supergravity, that 
all the soft trilinear parameters are proportional to their corresponding couplings
in the superpotential (for a review, see e.g. Ref.~\cite{Brignole:1997dp}), one can write
\bea
T^{e}_{ij} &=& A^{e}_{ij} Y^{e}_{ij}\ , \;\;
T^{d}_{ij} = A^{d}_{ij} Y^{d}_{ij}\ , \;\;
T^{u}_{ij} = A^{u}_{ij} Y^{u}_{ij}\ ,
\label{tyukawa}
\\
T^{\nu}_{ij} &=& A^{\nu}_{ij} Y^{\nu}_{ij}\ , \;\;
T^{\lambda}_i= A^{\lambda}_i\lambda_i\ , \;\;
T^{\kappa}_{ijk}= A^{\kappa}_{ijk} \kappa_{ijk}\ ,
\label{tmunu}
\eea
where $A\sim$ TeV, and the summation convention on repeated indexes does not apply.

\subsection{The 
Scalar Potential}
\label{sec:higgs potential}
In addition to terms from $\mathcal{L}_{\text{soft}}$, the tree-level neutral scalar potential  receives the usual $D$- and $F$-term contributions, $V^0=V_{\text{soft}} + V_F + V_D$, with
\bea
V_{\text{soft}}  =&&
\left(
T^{\nu}_{ij} \, H_u^0\,  \widetilde \nu_{iL} \, \widetilde \nu_{jR}^* 
- T^{\lambda}_{i}\, \widetilde \nu_{iR}^*\, H_d^0  H_u^0
+ \frac{1}{3} T^{\kappa}_{ijk} \, \widetilde \nu_{iR}^* 
\widetilde \nu_{jR}^* 
\widetilde \nu_{kR}^*\
+
\text{h.c.} \right)
\nonumber\\
&+&
m_{\widetilde{L}_{ij}}^2\widetilde{\nu}_{iL}^* \widetilde\nu_{jL}
+
m_{\widetilde{\nu}_{ij}}^2 \widetilde{\nu}_{iR}^* \widetilde\nu_{jR}
+
m_{H_d}^2 {H^0_d}^* H^0_d + m_{H_u}^2 {H^0_u}^* H^0_u,
\label{akappa}
\\
\nonumber
\\
%
V_{F}  =&&
 \lambda_{j}\lambda_{j}^{*}H^0_{d}H_d^0{^{^*}}H^0_{u}H_u^0{^{^*}}
 +
\lambda_{i}\lambda_{j}^{*}\tilde{\nu}^{*}_{iR}\tilde{\nu}_{jR}H^0_{d}H_d^0{^*}
 +
\lambda_{i}\lambda_{j}^*
\tilde{\nu}^{*}_{iR}\tilde{\nu}_{jR}  H^0_{u}H_u^0{^*}   
\nonumber\\                                              
&+&
\kappa_{ijk}\kappa_{ljm}^{*}\tilde{\nu}^*_{iR}\tilde{\nu}_{lR}
                                   \tilde{\nu}^*_{kR}\tilde{\nu}_{mR}
- \left(\kappa_{ijk}\lambda_{j}^{*}\tilde{\nu}^{*}_{iR}\tilde{\nu}^{*}_{kR} H_d^{0*}H_u^{0*}                                      
 -Y^{\nu}_{ij}\kappa_{ljk}^{*}\tilde{\nu}_{iL}\tilde{\nu}_{lR}\tilde{\nu}_{kR}H^0_{u}
 \right.
 \nonumber\\
 &+&
 \left.
 Y^{\nu}_{ij}\lambda_{j}^{*}\tilde{\nu}_{iL} H_d^{0*}H_{u}^{0*}H^0_{u}
+{Y^{\nu}_{ij}}^{*}\lambda_{k} \tilde{\nu}_{iL}^{*}\tilde{\nu}_{jR}\tilde{\nu}_{kR}^* H^0_{d}
 + \text{h.c.}\right) 
\nonumber \\
 &+& 
Y^{\nu}_{ij}{Y^{\nu}_{ik}}^* \tilde{\nu}^{*}_{jR}
\tilde{\nu}_{kR}H^0_{u}H_u^0{^*}                                                
 +
Y^{\nu}_{ij}{Y^{\nu}_{lk}}^*\tilde{\nu}_{iL}\tilde{\nu}_{lL}^{*}\tilde{\nu}_{jR}^{*}
                                  \tilde{\nu}_{kR}  
 +
Y^{\nu}_{ji}{Y^{\nu}_{ki}}^*\tilde{\nu}_{jL}\tilde{\nu}_{kL}^* H^0_{u}H_u^{0*},
\\
\nonumber
\\
V_D  =&&
\frac{1}{8}\left(g^{2}+g'^{2}\right)\left(\widetilde\nu_{iL}\widetilde{\nu}_{iL}^* 
+H^0_d {H^0_d}^* - H^0_u {H^0_u}^* \right)^{2}.
\label{dterms}
\eea
The scale of the soft terms in Eq.~(\ref{akappa}) is in the ballpark of {one} TeV, and they
induce the EWSB in the $\mn$.
The eight minimization conditions with respect to Higgses and sneutrinos, with the choice of CP conservation,\footnote{The $\mu\nu$SSM with spontaneous  CP violation was studied in Ref.~\cite{Fidalgo:2009dm}.} can be found in Refs.~\cite{Escudero:2008jg,Ghosh:2017yeh}.
For neutral Higgses, 
and right 
and  left 
sneutrinos defined as
\bea
H_d^0 
=
\frac{1}{\sqrt 2} \left(H_{d}^\mathcal{R} + v_d + i\ H_{d}^\mathcal{I}\right),\quad
H^0_u 
=
\frac{1}{\sqrt 2} \left(H_{u}^\mathcal{R}  + v_u +i\ H_{u}^\mathcal{I}\right),  
\\
\label{vevu}
\nonumber 
\\
\widetilde{\nu}_{iR} 
=
      \frac{1}{\sqrt 2} \left(\widetilde{\nu}^{\mathcal{R}}_{iR}+ v_{iR} + i\ \widetilde{\nu}^{\mathcal{I}}_{iR}\right),  
      \quad
  \widetilde{\nu}_{iL} 
  =
  \frac{1}{\sqrt 2} \left(\widetilde{\nu}_{iL}^\mathcal{R} 
  + v_{iL} +i\ \widetilde{\nu}_{iL}^\mathcal{I}\right),
\label{vevnu}
\eea
the following VEVs are developed: 
\begin{eqnarray}
\langle H_{d}^0\rangle = \frac{v_{d}}{\sqrt 2},\quad 
\langle H_{u}^0\rangle = \frac{v_{u}}{\sqrt 2},\quad 
\langle \widetilde \nu_{iR}\rangle = \frac{v_{iR}}{\sqrt 2},\quad 
\langle \widetilde \nu_{iL}\rangle = \frac{v_{iL}}{\sqrt 2},
\end{eqnarray}
with $v_{iR}\sim$ TeV {whereas 
$v_{iL}\sim 10^{-4}$ GeV}. 
The small values of $v_{iL}$ 
are because of
the proportional contributions to 
$Y^{\nu}$ appearing in the $v_{iL}$ minimization equations~\cite{LopezFogliani:2005yw,Escudero:2008jg,Ghosh:2017yeh}.
These contributions enter through $V_F$ and 
$V_{\text{soft}}$ (assuming 
$T^{\nu}=A^{\nu}Y^{\nu}$ as in Eq.~(\ref{tmunu}))
and 
are small
due to the generalized EW seesaw discussed in the Introduction that
determines $Y^{\nu}\lsim 10^{-6}$.
An easy estimation gives $v_{iL}\lsim m_{{\mathcal{D}_i}}$, with
\bea
m_{{\mathcal{D}_{ij}}}= Y^{\nu}_{ij} 
\frac{v_u}{\sqrt 2},
\label{diracmass}
\eea 
the Dirac masses for neutrinos.

%
%
%
As can be {straightforwardly} deduced from the fifth and sixth terms of $W$ in Eq.~(\ref{superpotential}), the effective $\mu$-term and Majorana masses are of the order TeV and given respectively by:
\bea
\mu= \la_i \frac{v_{iR}}{\sqrt 2},\: \: \: \: \: \: \: \: \; \: \: \: \: \: 
{\mathcal M}_{ij} = {2}\kappa_{ijk} \frac{v_{kR}}{\sqrt 2}.
\label{eq:mu}    
\eea


\subsection{The Parameter Space}

Given the structure of the scalar potential,
the free parameters in the neutral scalar sector of the $\mn$
at the low scale 
$M_{EWSB}= \sqrt{m_{\tilde t_l} m_{\tilde t_h}}$ 
are therefore:
$\lambda_i$, $\kappa_{ijk}$, $Y^{\nu}_{ij}$, $m_{H_{d}}^{2}$, $m_{H_{u}}^{2}$,  
$m_{\widetilde{\nu}_{ij}}^2$,
$m_{\widetilde{L}_{ij}}^2$, 
$T^{\lambda}_i$, $T^{\kappa}_{ijk}$ and $T^{\nu}_{ij}$.
Using diagonal sfermion mass matrices, in order to avoid the
strong upper bounds upon the intergenerational scalar mixing (see e.g. Ref.~\cite{Gabbiani:1996hi}), from the eight minimization conditions with respect to $v_d$, $v_u$,
$v_{iR}$ and $v_{iL}$
one can eliminate
the above
soft masses 
in favor
of the VEVs.
In addition, using 
$\tan\beta\equiv {v_u}/{v_d}$ 
and the SM Higgs VEV, $v^2 = v_d^2 + v_u^2 + \sum_i v^2_{iL}={4 m_Z^2}/{(g^2 + g'^2)}\approx$ (246 GeV)$^2$
with the electroweak gauge couplings estimated at the $m_Z$ scale by
$e=g\sin\theta_W=g'\cos\theta_W$, 
one can determine the SUSY Higgs 
VEVs, $v_d$ and $v_u$. 
Since $v_{iL} \ll v_d, v_u$, one has 
$v_d\approx v/\sqrt{\tan^2\beta+1}$.
Besides, one can use diagonal neutrino Yukawa couplings, since data of neutrino physics can
easily be reproduced at tree level in the $\mn$ with such structure, as we will discuss in 
Subsec.~\ref{lightn}.
Finally, assuming for simplicity that the 
off-diagonal elements of $\kappa_{ijk}$
and soft trilinear parameters $T$ vanish,
we are left with   
the following set of variables as independent parameters in the neutral scalar sector:
\bea
\lambda_i, \, \kappa_{i}, \, Y^{\nu}_{i}, \, \tan\beta, \, v_{iL}, \, v_{iR}, \, T^{\lambda}_i, 
\, T^{\kappa}_{i}, \, T^{\nu}_{i}.
\label{softfree}
\eea
where $\kappa_i\equiv\kappa_{iii}$,
$Y^{\nu}_i\equiv Y^{\nu}_{ii}$,
$T^{\nu}_i\equiv T^{\nu}_{ii}$
and
$T^{\kappa}_i\equiv T^{\kappa}_{iii}$.
Note that now the Majorana mass matrix in Eq.~(\ref{eq:mu}) is diagonal, with the non-vanishing entries given by
\bea
{\mathcal M}_{i}
={2}\kappa_{i} \frac{v_{iR}}{\sqrt 2},
\label{majorana2}
\eea
where the summation convention on repeated indexes does not apply.

The rest of (soft) parameters of the model, namely the following gaugino masses,
scalar masses, 
and trilinear parameters:
%
\bea
M_1, \, M_2,\, M_3, \, m_{\tilde Q_{i}},\, 
m_{\tilde u_{i}}, \, m_{\tilde d_{i}}, \,
m_{\tilde e_{i}}, \,
T^u_{i}, \, T^d_{i}, \, T^e_{i},
\label{freeparameterssoft}
\eea
are also taken as free parameters and specified at low scale.

\subsection{The Number of Right-Handed Neutrinos}
\label{thenumber}

{In our previous discussions, we have assumed that the number of RH neutrinos is three.
This is a natural value in the sense that replicates what happens with the other fermions of the SM. Nevertheless, {we would like to point out 
that} this number is in principle a free parameter. 
The RH neutrino is the only singlet of the SM and therefore its way of arising from a more fundamental theory does not have to be the same as for the other particles. 
Besides, {being} a singlet it does not contribute to gauge anomalies and therefore we can have any number of them. 

From the model-building viewpoint only one RH neutrino is sufficient to generate dynamically the $\mu$ term~\cite{LopezFogliani:2005yw,Bartl:2009an,Biekotter:2017xmf,Lara:2018rwv,Lara:2018zvf}, reproducing neutrino physics thanks to loop corrections~\cite{Bartl:2009an}.
Two RH neutrinos are sufficient to reproduce at tree level the neutrino mass differences and mixing angles, and it is also the minimal case with the capability of giving 
spontaneous CP violation in the neutrino sector \cite{Fidalgo:2009dm}.}
Moreover, more than three RH neutrinos can also be used, without spoiling any of the useful properties present in the case of three families.
As will be discussed in Subsec.~\ref{dmneutrinos},
some of these RH neutrinos can also behave as sterile neutrinos being candidates for DM. 
In what follows we 
will continue {working with} three RH neutrino superfields.


\subsection{Proton Decay and Naturalness Problems}
\label{proton}

We mentioned in the Introduction that the new couplings introduced in the $\mn$ are harmless for
proton decay. 
However, in purity one could still argue that RPC is mandatory to avoid the couplings
$\lambda'_{ijk}\hat L_i \hat Q_{j} \hat d^{c}_{k}$ 
and $\lambda''_{ijk}\hat d^c_{i} \hat d^c_{j} \hat u^c_{k}$.
Our viewpoint is that this imposition is clearly too stringent.
The choice of RPC
is {\it ad hoc}, since one can use other $Z_N$ discrete symmetries to forbid only the couplings
$\lambda''_{ijk}$, which is sufficient to avoid fast proton decay. This is e.g. the case of $Z_3$ Baryon-parity~\cite{Ibanez:1991hv,Ibanez:1991pr} which also prohibits dimension-5 proton decay operators, unlike $R$-parity.
Besides, this strategy seems reasonable if one expects all discrete symmetries to arise from the breaking of gauge symmetries of the underlying unified theory~\cite{Krauss:1988zc}, 
since
Baryon-parity and 
$R$-parity are the only two generalized parities which are `discrete gauge' anomaly free~\cite{Ibanez:1991hv,Ibanez:1991pr}.
Discrete gauge symmetries are also not violated~\cite{Krauss:1988zc} by potentially dangerous
quantum gravity effects~\cite{Gilbert:1989nq}. 

We would like also to remark that,
given the relevance of string theory as a possible underlying unified theory, a robust argument in favor of the above mechanism is that in string compactifications, such as e.g. orbifolds,
the matter superfields have
several extra $U(1)$ charges broken spontaneously at high energy by a Fayet-Iliopoulos D-term, and as a consequence residual $Z_N$ symmetries 
are left in the low-energy theory.
As pointed out in Ref.~\cite{Escudero:2008jg},
the same result can be obtained by the complementary mechanism 
that stringy selection rules can naturally forbid the $\lambda''_{ijk}$ couplings discussed above. This is because matter superfields are located in general in different sectors of the compact space, in such a way that
some RPV couplings can be forbidden but others are allowed~\cite{Casas:1987us,Casas:1988vk}.
In fact, we expect naturally the {presence of the lepton-number violating couplings $\lambda_{ijk} \hat L_i \hat L_j \hat e^c_k$ and $\lambda'_{ijk}\hat L_i \hat Q_{j} \hat d^{c}_{k}$ in the superpotential} of the $\mn$, as will be discussed in Subsec.~\ref{trilinearl}.




It is tempting to connect also string theory with the fact that the superpotential of the $\mn$ has a $Z_3$ discrete symmetry.
Thanks to it, the presence of the dangerous linear (tadpole) terms $t_{i} \hat\nu_{i}^c$ is forbidden, as well as that of the bilinear (mass) terms
$\mu\hat H_u \hat H_d$, $\mu_i \hat H_u \hat L_i$ and 
${\mathcal M}_{ij} \hat\nu_{i}^c\hat\nu_{j}^c$. The latter bilinears would reintroduce 
the $\mu$-problem and additional naturalness problems. 
In fact, a $Z_3$ symmetry is what one would expect from a high-energy theory where
the low-energy modes should be massless and the massive modes of the order of the high-energy scale.
As pointed out in Ref.~\cite{Lopez-Fogliani:2017qzj}, this is precisely the situation in 
string constructions,
where the massive modes have huge masses of the order of the string scale and the massless ones have only trilinear terms at the renormalizable level. Thus one ends up with an accidental $Z_3$ symmetry in the low-energy theory. The potential cosmological domain wall problem induced by this discrete symmetry
will be discussed in Subsec.~\ref{domain}.

Another interesting strategy to forbid the above dangerous operators in the $\mn$ is through the
{addition of an extra $U(1)'$ gauge symmetry}.
We will briefly review this matter in Sec.~\ref{gauge}.

\subsection{Comparison with Other Models}
\label{comparison}

Before continuing to analyze the characteristics of the $\mn$, it is worth stopping for a moment 
and comparing this model with others that are found in the literature.
We can use for the comparison the superpotential of the $\mn$ in Eq.~(\ref{superpotential}).
As discussed in the Introduction, 
the main difference with respect to the MSSM is that neutrino physics is included in the
model and no explicit 
$\mu$-term
$\epsilon_{ab}\mu\hat H^b_u \hat H^a_d$ is necessary in $W$, arising dynamically after EWSB.

A similar comment applies to the comparison of the $\mn$ with the 
bilinear $R$-parity violating model (BRPV)~\cite{Hall:1983id,Lee:1984kr,Lee:1984tn,Dawson:1985vr}, an extension of the MSSM where explicit lepton-number violating (mass) terms
$\epsilon_{ab}\mu_i \hat H^b_u \hat L^a_i$ are added.
The main motivation for the latter terms is to induce neutrino masses through the mixing of LH neutrinos 
with the
neutralinos (because of RPV), without the necessity of including RH neutrinos in the model.
However, the structure of the neutral fermion mass matrix is such that only one neutrino acquires mass
at tree level, while the other two get them through loop corrections. Another drawback,
as discussed in the previous subsection, is that the
$\mu$-problem~\cite{Kim:1983dt} is augmented with the three new bilinear terms $\mu_i$ which should be of the order of
$10^{-4}$ GeV to reproduce correct neutrino masses.
In the $\mn$,
those terms are generated dynamically from the Dirac Yukawa couplings for neutrinos
in Eq.~(\ref{superpotential}):
\beq
\mu_i=Y^{\nu}_{ij}\frac{v_{jR}}{\sqrt 2}.\label{bilinearterm}
\eeq

Other well-known RPV models are those involving conventional trilinear 
couplings (TRPV)~\cite{Weinberg:1981wj,Sakai:1981pk}.
There are two possibilities, to add to the superpotential of the MSSM the lepton-number violating couplings
$\epsilon_{ab}(\lambda_{ijk} \hat L^a_i \hat L^b_j \hat e^c_k + \lambda'_{ijk}\hat L^a_i \hat Q^b_{j} \hat d^{c}_{k})$, or the baryon-number violating couplings 
$\lambda''_{ijk}\hat d^c_{i} \hat d^c_{j} \hat u^c_{k}$.
In the case of the former model, neutrino masses can arise through loop corrections, but
both types of models do not attempt to solve the $\mu$-problem.
For a review about the RPV models BRPV and TRPV,
see e.g. Ref.~\cite{Barbier:2004ez}.
{As mentioned in the previous subsection}, the $\lambda_{ijk}$ and $\lambda'_{ijk}$ couplings
are expected to be naturally present in
the superpotential of the $\mn$.

Concerning models proposing solutions to the $\mu$-problem, an elegant one is the
RPC next-to-minimal supersymmetric standard model (NMSSM) (for a review, see e.g. Refs.~\cite{Maniatis:2009re,Ellwanger:2009dp}). There, an extra singlet superfield $\hat N$ is introduced in $W$ coupled to Higgses and to itself: $-\epsilon_{ab}\lambda \hat N \hat H^b_u \hat H^a_d + \frac{1}{3}\kappa \hat N\hat N\hat N$.
The differences with the $\mn$ are obvious. An extra field, different from the RH neutrino, has to be added to the spectrum, and no attempt to solve the $\nu$-problem is made.
We can obtain the NMSSM in the limit of the $\mn$ when Dirac Yukawa couplings for neutrinos are
set to zero, $Y^{\nu}_{ij}=0$.
Another interesting model~\cite{Kitano:1999qb} is obtained by extending the NMSSM with couplings between the singlet and the RH neutrino superfields $\kappa_{ij} \hat N\hat \nu^c_i\hat \nu^c_j$.
Here Majorana masses for RH neutrinos are generated through the VEV of the singlet $N$. 
In this model, in addition to the neutralino also the right sneutrino can be a viable DM 
candidate~\cite{Cerdeno:2008ep}. 


\section{The Spectrum}
\label{sec:spectrum}

This section is devoted to review the spectrum of the $\mn$, paying special attention 
to the sectors where the recent experimental results impose strong constraints, i.e. neutrino and Higgs sectors.
Besides, unlike others the structure of these two sectors is very different from the MSSM ones. In addition, reproducing neutrino data
is an important asset of the model as described in the Introduction.

As shown in the previous section, 
the distinctive couplings of the $\mn$, 
$Y^{\nu}_{ij}$, $\lambda_i$ and $\kappa_{ijk}$, contribute to the neutral scalar potential generating as a consequence VEVs for Higgses, and right and left sneutrinos. 
A detailed analysis of the Higgs sector was first performed in Ref.~\cite{Escudero:2008jg},
finding viable regions that avoid false minima and tachyons, as well as fulfill the Landau pole constraint.
The complete one-loop renormalization of the neutral scalar sector of the $\mn$
was carried out in Refs.~\cite{Biekotter:2017xmf,Biekotter:2019gtq}.
Because of the new couplings and VEVs, {\bf all fields in the spectrum with the same color, electric charge and spin mix together}. This is similar to the MSSM, where the couplings and Higgs VEVs determine the mixing of bino, wino and higgsinos, producing the four neutralino states. However, in the MSSM the RPC prevents mixing of states with different 
$R$-parity quantum numbers, such as e.g. neutrinos with neutralinos, or sneutrinos with Higgses. This is not the case of the $\mn$, where RPV allows this kind of mixing, and, as a consequence, a rich phenomenology. 

The associated mass matrices were studied in detail in
Refs.~\cite{Escudero:2008jg,Ghosh:2008yh,Bartl:2009an,Ghosh:2017yeh}. 
In particular, Appendix B of Ref.~\cite{Ghosh:2017yeh} contains all the matrices and is very useful for the discussion that follows. 
Summarizing the results, there are the following relevant states:
eight neutral scalars and seven neutral pseudoscalars (Higgses-sneutrinos), 
seven charged scalars (charged Higgses-sleptons),
five charged fermions (charged leptons-charginos),
and 
ten neutral fermions (neutrinos-neutralinos).

\subsection{The Neutrino Sector}
\label{neutralinos}


LH neutrinos mix with MSSM neutralinos and RH neutrinos, thus
the {\bf neutral fermions} have the flavor composition 
$({\chi^{0}})^T=(
{(\nu_{iL})^c}^*,
\widetilde B^0,
\widetilde W^{0},\widetilde H_{d}^0,\widetilde H_{u}^0,\nu^*_{iR})$,
%
%
and one obtains the mass terms, 
$-\frac{1}{2} ({\chi^{0}})^T 
{m}_{\nu} 
\chi^0 + \mathrm{h.c.}$,
with ${m}_{\nu}$ the $10\times 10$ (symmetric) `neutrino' mass matrix: 
\begin{equation}
{m}_{\nu} 
=\left(\begin{array}{cc}
 0_{3\times3} & m\\
m^{T} & {\mathcal M}_{7\times 7} \end{array}\right).
\label{matrizse}
\end{equation}
In this matrix,
$m$ is a $3\times 7$ submatrix containing the mixing of LH neutrinos with MSSM neutralinos and RH neutrinos.
The submatrix 
${\mathcal M}_{7\times 7}$
contains the mixing of MSSM neutralinos with RH neutrinos, in addition
of the mixing between MSSM neutralinos themselves and RH neutrinos themselves.

It is relevant to note that ${m}_{\nu}$  has the structure of a generalized EW seesaw. 
Note in this respect that all the entries of ${\mathcal M}_{7\times 7}$
are of the order of TeV. This is because 
the low-energy bino and wino soft masses, $M_1$ and $M_2$, are of that order, as well as the $\mu$ term, and the mixing of neutralinos with RH neutrinos which is determined by $\lambda_i v_u$,
$\lambda_i v_d$. In addition, the self mixing of the RH neutrinos is determined by
the Majorana masses of Eq.~(\ref{eq:mu}) which are $\sim$ TeV.
On the contrary, the entries of the matrix $m$ are much smaller:
the Dirac neutrino masses are $\lsim 10^{-4}$ GeV as discussed
in Eq.~(\ref{diracmass}), and the other entries 
are of similar order
being determined by $g v_{iL}$, $g' v_{iL}$ 
and $Y^{\nu}_{ij} v_{jR}$.

\subsubsection{Light Neutrinos}
\label{lightn}

The above generalized EW seesaw
produces {three light neutral fermions dominated by the 
$\nu_{iL}$
flavor composition}. 
As shown in Refs.~\cite{LopezFogliani:2005yw,Escudero:2008jg,Ghosh:2008yh,Bartl:2009an,Fidalgo:2009dm,Ghosh:2010zi,Liebler:2011tp},
data on neutrino physics~\cite{Capozzi:2017ipn,deSalas:2017kay,deSalas:2018bym,Esteban:2018azc} can easily be reproduced at tree level,
even with diagonal Yukawa couplings~\cite{Ghosh:2008yh,Fidalgo:2009dm}, i.e.
$Y^{\nu}_{ii}=Y^{\nu}_{i}$ and vanishing otherwise.
%
A simplified formula 
for the effective mixing mass matrix of the 
light neutrinos is~\cite{Fidalgo:2009dm}:
\begin{eqnarray}
\label{Limit no mixing Higgsinos gauginos}
(m_{\nu})_{ij} 
=&&
\frac{m_{{\mathcal{D}_i}} m_{{\mathcal{D}_j}} }
{3{\mathcal{M}}}
                   (1-3 \delta_{ij})
                   -\frac{(v_{iL}/\sqrt 2)(v_{jL}/\sqrt 2)}
                   {2M^{\text{eff}}}
\nonumber\\
&-&
\frac{
m_{{\mathcal{D}_i}}m_{{\mathcal{D}_j}}
}
{2M^{\text{eff}}}
\frac{1}{3\lambda\tan\beta}
\left(
\frac{v_{iL}/\sqrt 2}{m_{{\mathcal{D}_i}}}
   +
   \frac{v_{jL}/\sqrt 2}{m_{{\mathcal{D}_j}}}
   + \frac{1}{3\lambda\tan\beta}
   \right),
\label{neutrinoph}
  \end{eqnarray}     
where 
\begin{eqnarray}
\label{effectivegauginomass}
 M^{\text{eff}}\equiv M -
 \frac{\left(v/\sqrt 2\right)^2
 }
 {2\mu
 \left({\mathcal{M}} \frac{v_R}{\sqrt 2}+ 2\lambda 
 \left(\frac{v}{\sqrt 2}\right)^2 
 \frac{\tan\beta}{1+\tan^2\beta}\right)
        }
        \left[2{\mathcal{M}} 
        \frac{v_R}{\sqrt 2} 
        \frac{\tan\beta}{1+\tan^2\beta}
        +\lambda \left(\frac{{v}}{\sqrt 2}\right)^2\right]
        \label{first}
\end{eqnarray} 
with
\begin{eqnarray}
\label{effectivegauginomass2}
\frac{1}{M} = \frac{g'^2}{M_1} + \frac{g^2}{M_2}.
\label{gauginom}
\end{eqnarray} 
Here
we have assumed universal $\lambda_i = \lambda$, $v_{iR}= v_{R}$, and
$\kappa_i=\kappa$.
In this case, the Dirac masses for neutrinos of Eq.~(\ref{diracmass}), the $\mu$-term of 
Eq.~(\ref{eq:mu}) and the three non-vanishing Majorana masses 
${\mathcal M}_i={\mathcal M}$ of Eq.~(\ref{majorana2}), are given by
\bea
m_{{\mathcal{D}_i}}\equiv Y^{\nu}_{i} \frac{v_u}{\sqrt 2}, \: \: \: \: \: \: \: \: \: \: \: \: \: \: \:
\mu=3\la \frac{v_{R}}{\sqrt 2}, \: \: \: \: \: \: \: \: \: \: \: \: \: \: \:
{\mathcal M}
={2}\kappa \frac{v_{R}}{\sqrt 2}.
\label{mu2}    
\eea
The first term of Eq.~(\ref{Limit no mixing Higgsinos gauginos}) is generated
through the mixing 
of $\nu_{iL}$ with 
$\nu_{iR}$-Higgsinos, and the other two
also include the mixing with gauginos.
{These are the so-called $\nu_{R}$-Higgsino seesaw and gaugino seesaw, 
respectively~\cite{Fidalgo:2009dm}.}
We are then left in general with the following subset of variables of
Eqs.~(\ref{softfree}) and~(\ref{freeparameterssoft})
as independent parameters in the neutrino sector:
\bea
\lambda_i,\, \kappa_i,\, Y^{\nu}_i, \tan\beta, \, v_{iL}, \, v_{iR}, \, M_1, \, M_2.
\label{freeparametersn}
\eea
In order to efficiently scan the model in phenomenological analyses, it is useful to reduce the number of parameters to be used as we will discuss in Subsec.~\ref{leftsneutrinos}. Thus, inspired by GUTs one can assume e.g. the low-energy relation, 
$M_2= (\alpha_2/\alpha_1) M_1\simeq 2 M_1$, 
where $g_2=g$ and 
$g_1=\sqrt{5/3}\ g'$,
and therefore the parameters $M_1$ and $M_2$ can be substituted in the computation by $M$ as the relevant parameter.

Under several assumptions, the formula for $(m_{\nu})_{ij}$ can be further simplified.
Notice first that the third term is inversely proportional to $\tan\beta$,
and therefore negligible in the limit of large or even moderate $\tan\beta$ provided that 
$\lambda$ is not too small. Besides,
the first piece inside the brackets in the second term of Eq.~(\ref{first}) 
is also negligible in this limit, and for typical values of the parameters involved in the seesaw also the second piece, thus
$M^{\text{eff}}\sim M$.
Under these assumptions, the second
term for $(m_{\nu})_{ij}$ 
is generated only through the mixing of LH neutrinos with gauginos. Therefore, we arrive to a very simple formula where only the first two terms survive with $M^{\text{eff}}= M$ in Eq.~(\ref{neutrinoph}):
\begin{eqnarray}
\label{Limit no mixing Higgsinos gauginos2}
(m_{\nu})_{ij} 
=
\frac{m_{{\mathcal{D}_i}} m_{{\mathcal{D}_j}} }
{3{\mathcal{M}}}
                   (1-3 \delta_{ij})
                   -\frac{(v_{iL}/\sqrt 2)(v_{jL}/\sqrt 2)}
                   {2M}.
\label{neutrinoph2}
  \end{eqnarray}     
This expression can be used to understand easily the
seesaw mechanism in the $\mn$ in a qualitative way.
From this discussion,
it is clear that $Y^{\nu}_i$, $v_{iL}$ and $M$ are crucial parameters to determine the neutrino 
physics. 
For typical values as those discussed in the previous section,
$Y^{\nu}_i \lsim 10^{-6}$, $v_{iL}\sim 10^{-4}$ GeV and $M\sim$ TeV, neutrino masses $\lsim 0.1$ eV as expected, can easily be reproduced.\footnote{
Note that if we had included 
$m^{2}_{H_d \tilde L_{iL}} {H_d^a}^*  \tilde L_{iL}^a + \text{h.c.}$ in $V_{\text{soft}}$ of Eq.~(\ref{akappa}), we would have obtained
$v_{iL}\sim$ TeV. 
Thus,
the second term of Eq.~(\ref{neutrinoph2}) would have given rise to LH neutrinos $\sim$ TeV.
We are assuming therefore that this type of soft masses are not present in our Lagrangian or that they are negligible.
A similar destabilization of $v_{iL}$ would arise
with $T^{\nu}\sim$ TeV in $V_{\text{soft}}$. The assumption $T^{\nu}=A^{\nu}Y^{\nu}$ of
Eq.~(\ref{tmunu})
solves this problem.
As discussed in detail in the Appendix A of Ref.~\cite{Ghosh:2017yeh},
these two assumptions about the structure of the soft terms 
are reliable in the framework of the current studies of SUSY and supergravity.
}

Actually, it is possible to go further establishing qualitatively what regions of the parameter space are
the best in order to obtain correct neutrino masses and mixing angles. 
In particular, {one} can determine natural hierarchies among 
neutrino Yukawas, and among left sneutrino VEVs.
Considering the normal ordering for the neutrino mass 
spectrum,
representative solutions for neutrino physics using diagonal neutrino Yukawas in this scenario are summarized below~\cite{Kpatcha:2019gmq}. Note that these solutions take advantage of the dominance of the gaugino seesaw for some of the three neutrino families.

1) $M<0$, with $Y^{\nu}_1<Y^{\nu}_2, Y^{\nu}_3$, and $v_{1L} > v_{2L}, v_{3L}$.

\noindent 
As explained in Refs.~\cite{Fidalgo:2009dm,GomezVargas:2017}, a negative value for $M$ is useful in order to reproduce neutrino data with $Y^{\nu}_1$ the smallest Yukawa and $v_{1L}$ the largest VEV. 
Essentially, this is because 
a small tuning in 
Eq.~(\ref{Limit no mixing Higgsinos gauginos}) 
between the gaugino seesaw and the $\nu_{R}$-Higgsino seesaw is
necessary in order to obtain the correct mass of the first family.
Here the contribution of the gaugino seesaw is always the largest one. 
On the contrary, for the other two neutrino families, the contribution of the $\nu_{R}$-Higgsino seesaw is the most important one and that of the gaugino seesaw is less relevant for the tuning.

2) $M>0$, with $Y^{\nu}_3 < Y^{\nu}_1 < Y^{\nu}_2$, and $v_{1L}<v_{2L}\sim v_{3L}$.

\noindent 
In this case, it is easy to find solutions with the gaugino seesaw as the dominant one for the third family. Then, $v_{3L}$ determines the corresponding neutrino mass and $Y^{\nu}_3$ can be small.
On the other hand, the normal ordering for neutrinos determines that the first family dominates the lightest mass eigenstate implying that $Y^{\nu}_{1}< Y^{\nu}_{2}$ and $v_{1L} < v_{2L},v_{3L}$, with both $\nu_{R}$-Higgsino and gaugino seesaws contributing significantly to the masses of the first and second family. Taking also into account that the composition of these two families in the second mass eigenstate is similar, we expect $v_{2L} \sim v_{3L}$. 

3) $M>0$, with $Y^{\nu}_2 < Y^{\nu}_1 < Y^{\nu}_3$, and $v_{1L}<v_{2L}\sim v_{3L}$.

\noindent 
These solutions can be deduced from the previous ones {in 2)} interchanging the values of the third family, $Y^{\nu}_3$ and $v_{3L}$, with the corresponding ones of the second family, $Y^{\nu}_2$ 
and $v_{2L}$.
A small adjust in the parameters will lead again to a point in the parameter space satisfying neutrino data. This is clear from the fact that $\theta_{13}$ and $\theta_{12}$ are not going to be significantly altered, whilst $\theta_{23}$ may require a small tuning in the parameters. 
If the gaugino seesaw dominates for the second family, $v_{2L}$ determines the corresponding neutrino mass and $Y^{\nu}_2$ can be small. 

\vspace{0.20cm}

Concerning phases in the neutrino sector, 
it was pointed out in Ref.~\cite{Fidalgo:2009dm} that spontaneous CP violation is possible 
in the $\mn$ through complex Higgs and sneutrino VEVs, and that this is a source of {\bf Dirac and Majorana phases in the neutrino sector}. 
The continue improvements in neutrino experiments including the measurements of CP
violating phases~\cite{Abe:2019vii,Esteban:2020cvm} are crucial nowadays, and the $\mn$ seems to have
much to say about them.


\vspace{0.20cm}

\noindent
Let us finally point out
that all these results give a kind of answer to the question of 
{\bf why the mixing angles are so different in the quark and lepton sectors}. In the framework of the $\mn$, this is basically because no generalized seesaw exists for the quarks.

\subsubsection{$\mn$ Neutralinos}

As discussed in Eq.~(\ref{matrizse}),
the submatrix ${\mathcal M}_{7\times 7}$ contains the mixing of MSSM neutralinos and RH neutrinos. We denote these seven states as $\mn$ neutralinos.
This matrix is of the NMSSM type, apart form the small corrections proportional 
to $Y^{\nu}_{ij}$, and the fact that in the NMSSM there is only one singlet.
Different compositions of the eigenstates are now possible.
For example, if we take the values of the Majorana masses ${\mathcal M}_{ij}$ in 
Eq.~(\ref{eq:mu}) small compared to the soft gaugino masses $M_1$ and $M_2$, and
to the Higgsino masses $\mu$, the lightest $\mn$ neutralino is mainly a RH neutrino.
This composition of the LSP is genuine of the $\mn$ and hence, very interesting to study.
Another limit is to take small enough values for $M_1$, in such a way that one has a MSSM lightest neutralino almost bino-like. 

The phenomenology associated to these states, neutralinos$-$RH neutrinos, was studied
in Refs.~\cite{Ghosh:2008yh,Bartl:2009an,Fidalgo:2011ky,Ghosh:2012pq,Ghosh:2014rha,Ghosh:2014ida,Lara:2018zvf}, and we will discuss it in some detail in Subsec.~\ref{neutralino}.



\subsection{The Higgs Sector}

In the $\mn$, neutral doublet-like Higgses mix with left and right sneutrinos, thus the CP-even {\bf neutral scalars} have the
composition 
$S^T=(H_{d}^\mathcal{R},H_{u}^\mathcal{R},\tilde{\nu}^{\mathcal{R}}_{iR},
\tilde{\nu}_{jL}^\mathcal{R})$, and one obtains the mass terms,
$
-\frac{1}{2} S^T 
{m}^2_{h} S$,
with ${m}^2_{h}$ the 
$8\times 8$ (symmetric) `Higgs' mass matrix.
The same discussion applies to CP-odd neutral scalars, although in this case 
after rotating away the
pseudoscalar would be Goldstone boson, we are left with seven 
pseudoscalar
states. 
The $5\times 5$ Higgs-right sneutrino submatrix is almost decoupled from
the $3\times 3$ left sneutrino submatrix,
since the mixing occurs through terms
proportional to $Y^{\nu}_{ij}$ or $v_{iL}$, 
and these quantities are very small in order to satisfy neutrino data.
Besides, similar to the neutralino sector, the former $5\times 5$ submatrix is of the NMSSM type, apart from the small corrections proportional to 
$Y^{\nu}_{ij}$, and the fact that in the NMSSM there is only one singlet.
Thus, to accommodate the SM-like Higgs in the $\mn$, one can focus on the analysis of the Higgs-right sneutrino mass submatrix.
Upon diagonalization of the scalar submatrix, one obtains the
SM-like Higgs, the heavy doublet-like neutral Higgs, and three singlet-like states.
Similarly, upon diagonalization of the pseudoscalar submatrix,
and after rotating away the pseudoscalar would be Goldstone boson, we are left with the doublet-like neutral pseudoscalar, and three singlet-like pseudoscalar states. 
In what follows, we will review these states, including also the charged Higgs sector.


\subsubsection{The SM-like Higgs}

The accommodation of the SM-like Higgs boson discovered at the LHC is mandatory for any SUSY model. A recent analysis in the $\mn$ was carried out in Ref.~\cite{Kpatcha:2019qsz}.
Taking into account all the contributions, the mass of the {SM-like Higgs} 
in the $\mn$ {can in a very simplified way schematically be written as~\cite{Escudero:2008jg,Ghosh:2014ida}:
%
\bea
m^{2}_{h} = 
m^2_{0h} 
+  \Delta_{\text{mixing}} + \Delta_{\text{loop}},
\label{boundHiggs}
\eea
where  
\bea
m^{2}_{0h} =
m^2_Z \cos^2 2\beta + 
({v}/{\sqrt 2})^2\
{\bm \lambda}^2 \sin^2 2\beta\;
\label{boundHiggs2}
\eea
contains 
two terms, the first one characteristic of the MSSM and the second one of the $\mn$ with 
\begin{equation}
{\bm \la}\equiv \left({\sum_i \la^2_i}\right)^{1/2}=\sqrt{3}\ \la,
\label{lambda123}
\end{equation}
where the last equality is obtained if one assumes for simplicity universality of the parameters
$\lambda_i=\lambda$.
Note that $m^{2}_{0h}$ in Eq.~(\ref{boundHiggs}) 
corresponds to the absence of mixing of the SM-like Higgs with the other states in the mass squared matrix. 
$\Delta_{\text{mixing}}$ encodes those mixing effects
lowering (raising) the mass if it mixes with heavier (lighter) states, and $\Delta_{\text{loop}}$ refers to the radiative corrections.
In Refs.~\cite{Biekotter:2017xmf,Biekotter:2019gtq}, a full one-loop calculation of the corrections to the neutral scalar masses was performed. Supplemented by MSSM-type corrections at the two-loop level and 
beyond (taken over from the code {\tt FeynHiggs}~\cite{Heinemeyer:1998yj,Hahn:2009zz,Bahl:2018qog}) it was shown that the $\mu\nu$SSM can easily accomodate a SM-like Higgs boson at $\sim 125$~GeV, while simultaneously being in agreement with collider bounds and neutrino data. 
}

One can write $m^{2}_{0h}$ in a more
elucidate form 
as
\bea
m^{2}_{0h} = {m^2_Z}
\left\{
\left(\frac{1 - {\rm tan}^2\beta}{1 + {\rm tan}^2\beta}\right)^2  +  
\left(\frac{v/\sqrt 2}{m_Z}\right)^2\
{\bm \lambda}^2 
\left(\frac{{\rm 2\ tan\beta}}{1 + {\rm tan}^2\beta}\right)^2
\right\},
\label{boundHiggs1}
\eea
where the factor
$({v/\sqrt 2 m_Z})^2\approx 3.63$,
and 
we see straightforwardly that the second term 
grows with small tan$\beta$ and large 
${\bm \la}$.
In the case of the MSSM this term 
is absent, hence the maximum possible tree-level mass is about $m_Z$ for
$\tb \gg 1$ and, consequently, a contribution 
from loops is essential to reach the target of a SM-like Higgs in the mass region around 125 GeV. 
This contribution is basically determined by the soft parameters
$T^{u}_{3}, m_{\widetilde u_{3}}$ and $m_{\widetilde Q_{3}}$.
On the contrary, 
in the $\mu\nu$SSM one can reach this mass solely 
with the tree-level contribution for large 
values of ${\bm \la}$~\cite{Escudero:2008jg}.

\subsubsection{Right Sneutrino-like States}

From the scalar and pseudoscalar mass 
submatrices (see Appendix A of Ref.~\cite{Kpatcha:2019qsz}), one can easily deduce
that $\kappa_i$ and $T^{\kappa}_i$ 
are crucial parameters to determine the masses of the singlet-like states, originating from the self-interactions.
The remaining parameters $\lambda_i$ and $T^{\lambda}_i$ ($A^{\lambda}_i$ assuming the supergravity relation 
$T^{\lambda}_i= \lambda_i A^{\lambda}_i$ of 
Eq.~(\ref{tmunu}))
not only appear in the said interactions, but also control the mixing between the singlet and the doublet states and hence, contribute in determining the mass scale.
Note that the contributions of the parameters $T^{\nu}_i$ are negligible assuming $T^{\nu}_i=Y^{\nu}_i A^{\nu}_i$,
given the small values of neutrino Yukawas.
We conclude, taking also into account the discussion below Eq.~(\ref{boundHiggs1}), that the relevant independent low-energy parameters
in the Higgs-right sneutrino sector are the following subset of parameters of
Eqs.~(\ref{softfree}) and~(\ref{freeparameterssoft}):
\bea
\lambda_i, \, \kappa_i, \tan\beta, \, v_{iR}, \, T^{\kappa}_i,
\, T^{\lambda}_i,
T^{u}_3, \, m_{\widetilde u_{3}}, \,
m_{\widetilde Q_{3}}.
\label{freeparameters}
\eea


\begin{figure}[t!]
 \centering
 \includegraphics[width=0.8\linewidth, height= 0.35\textheight]{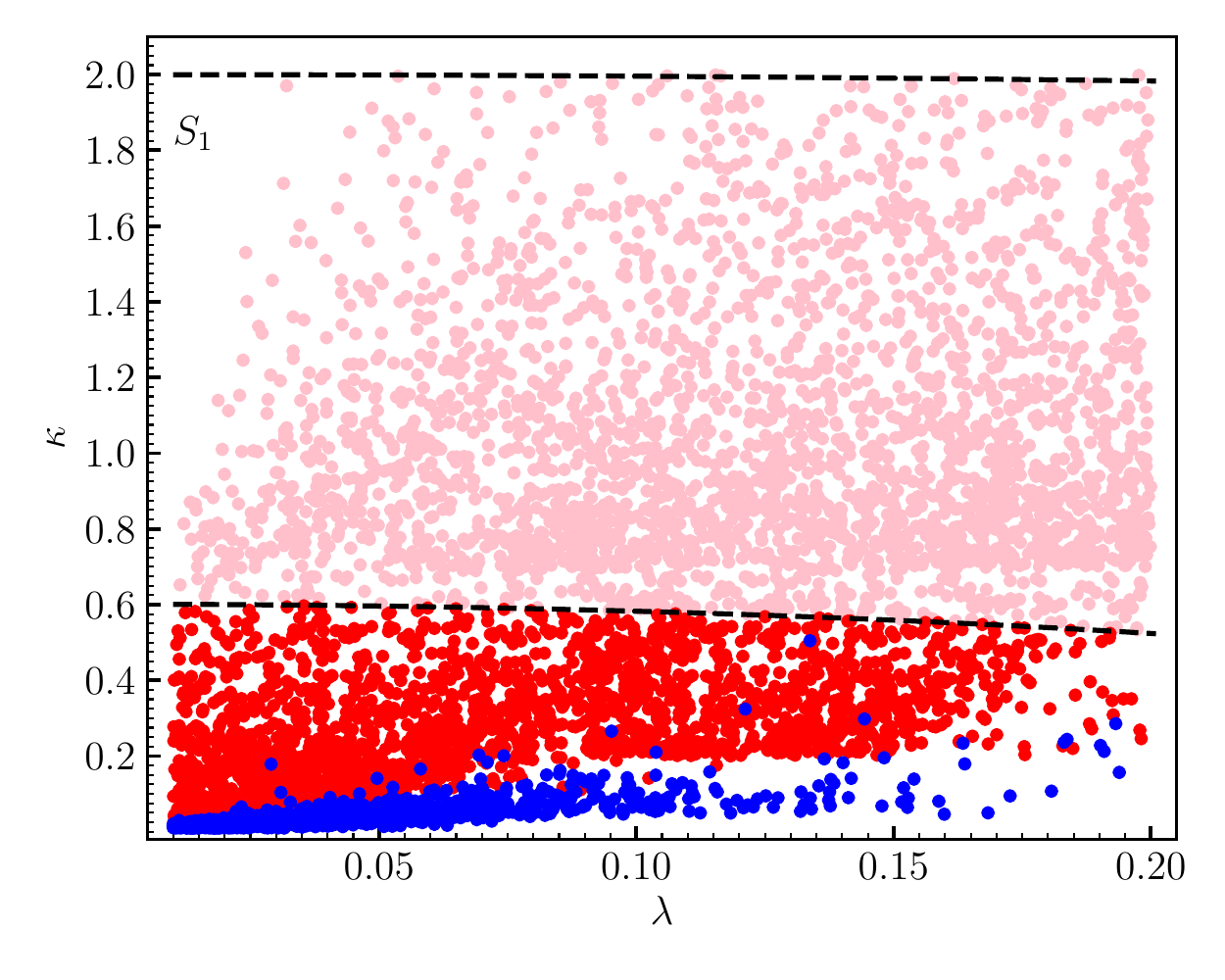}
  \caption{
 Viable points of the parameter space of the $\mn$ in the $\kappa-\lambda$ plane for the 
scan $S_1$ with $0.01\leq\lambda<0.2$, from Ref.~\cite{Kpatcha:2019qsz}.
The red and light-red (blue) colours represent cases where the SM-like Higgs is (is not) the lightest scalar.
All red and blue points below the lower black dashed line fulfill the perturbativity condition up to GUT scale, 
$2.77\ \kappa^2 + 2\ \bla^2\lsim 1$. 
Light-red points below the upper black dashed line fulfill the perturbativity condition up to 10 TeV,
$0.25\ \kappa^2 + 0.14\ \bla^2\lsim 1$.
A discussion about both conditions can be found in Ref.~\cite{Kpatcha:2019qsz}.
}
\label{S1Lambda-kappa}
\end{figure}

For the {right sneutrino-like states},
to obtain approximate analytical formulas for tree-level masses is only possible in
special regions of the parameter space of the $\mn$
(for a recent analysis see~\cite{Kpatcha:2019qsz} and references therein).
For example, for small $\lambda_i$ values (i.e. $\lsim$ 0.01) the singlet states are basically decoupled from the doublets, and one can write the 
right sneutrino masses for scalars and pseudoscalars as 
\bea
m_{\widetilde{\nu}^{\mathcal{R}}_{iR} 
}^2
=
 \left(\frac{T^{\kappa}_i}{\kappa_i} 
 +2{\mathcal M}_{i}\right)\frac{{\mathcal M}_{i}}{2},
 \: \: \: \: \: \: \: \: \; \: \: \: \: \: 
m_{\widetilde{\nu}^{\mathcal{I}}_{iR} 
}^2
=
-\frac{3}{2} \frac{T^{\kappa}_i}{\kappa_i} {\mathcal M}_{i},
\label{evenRR}
\eea
where in the case of supergravity we can use the relation 
$T^{\kappa}_i/\kappa_i= A^{\kappa}_i$.
Note that the parameters $\kappa_i, v_{iR}$ and $T^{\kappa}_i$ are the key ingredients to
determine the mass scale of the right sneutrino states~\cite{Escudero:2008jg,Ghosh:2008yh}.
Besides, in this limit of small $\lambda_i$ values, ${\mathcal M}_{i}$ coincide approximately with
the masses of the right-handed neutrinos, since they are decoupled from the other entries of the neutralino mass matrix:
\bea
m_{\nu_{iR}} = {\mathcal M}_{i}.
\label{neutrinoR}
\eea
With the sign convention ${\mathcal M}_{i}>0$, from the pseudoscalar mass formula of
Eq.~(\ref{evenRR}) one can deduce that
negative values for $T^{\kappa}_i$ (or $A^{\kappa}_i$) 
are necessary in order to avoid tachyonic pseudoscalars.
Then, from the scalar mass formula we can deduce that {\bf singlet scalars lighter than the SM-like Higgs can be obtained}. This result of light scalars (and pseudoscalars) in certain regions of the parameter space is generic, applying also to larger values of $\lambda_i$.
This was obtained in
Ref.~\cite{Kpatcha:2019qsz}, where a scan of the parameter space of the $\mn$ Higgs sector was carried out (see the discussion in Subsec.~\ref{higgsd} for more details).
As an example, we show in Fig.~\ref{S1Lambda-kappa} the result of the scan $S_1$ for the region of small/moderate values of $\lambda$ ($0.01\leq\lambda<0.2$). The results for larger values 
of $\lambda$ can also be found in that work, in particular for moderate/large values ($0.2\leq\lambda<0.5$) in scan $S_2$ and for
large values ($0.5\leq\lambda<1.2$) in scan $S_3$.



In Refs.~\cite{Escudero:2008jg,Bandyopadhyay:2010cu,Fidalgo:2011ky,Ghosh:2012pq,Ghosh:2014ida,Biekotter:2017xmf,Biekotter:2019gtq,Kpatcha:2019qsz},
it was found that large regions of the parameter space compatible with neutrino physics are viable, containing an interesting phenomenology related to the right sneutrino-Higgs sector that could be probed at the
LHC.
These are processes such as e.g. new two-body decays for the SM-like Higgs in the presence of light scalars, pseudoscalars and neutralinos. More details will be discussed in Subsec.~\ref{higgsd}.

\subsubsection{Left Sneutrinos}
\label{lhs}

The behaviour of the {left sneutrinos} is very different from the one of the right sneutrinos, since the former are tightly associated to neutrino physics.
As discussed before, the $3\times 3$ scalar and pseudoscalar left sneutrino submatrices are decoupled from the $5\times 5$ Higgs-right sneutrino sumatrices.
Besides, their off-diagonal entries
are negligible compared to the diagonal ones, since they
are suppressed by terms proportional 
to 
${Y^{\nu}_{ij}}^2$ 
and $v^2_{iL}$. As a consequence, the mass
squared eigenvalues correspond to the diagonal entries, and in this approximation both states also have degenerate masses.
Using the minimization equations for $v_{iL}$, one can write their tree-level 
values as~\cite{Escudero:2008jg,Ghosh:2008yh,Ghosh:2017yeh,Biekotter:2019gtq}
\bea
m_{\widetilde{\nu}^{\mathcal{R}}_{iL}}^2
=
m_{\widetilde{\nu}^{\mathcal{I}}_{iL}}^2
=
\frac{
m_{{\mathcal{D}_i}}}{v_{iL}/\sqrt 2}
\frac{v_{iR}}{\sqrt 2}
\left[\frac{-T^{\nu}_i}{Y^{\nu}_i}-
\frac{{\mathcal M}_{i}}{2}
+ \frac{\mu}{\tan\beta}+\lambda_i \frac{\left(v/\sqrt 2\right)^2}{v_{iR}/\sqrt 2}
\frac{\tan\beta}{1+\tan^2\beta}\right].
\label{evenLLL2}
\eea
Therefore, 
the 
\bea
T^{\nu}_i,
\label{tsneutrino}
\eea
are relevant parameters for the study of left sneutrino masses.
Note that the fourth term in Eq.~(\ref{evenLLL2}) 
can usually be neglected as long as $v_{iR} \gg v$ and/or $\lambda_i$ 
is small. Also, in the limit of moderate/large $\tan\beta$ (and not too large $\lambda_i$ compared to $\kappa_i$) one can neglect the third term compared to the second one.
{Given the sign convention above of positive Majorana masses ${\mathcal M}_{i}$, it is then convenient to use negative values for $T^{\nu}_i$
in order to avoid tachyonic states.

Now, from Eq.~(\ref{evenLLL2}) we see clearly why 
{\bf left sneutrinos are special} in the $\mn$ with respect to other SUSY models. 
Given 
that their masses are determined by the minimization equations with respect to
$v_{iL}$, they depend not only on left sneutrino VEVs but also on neutrino Yukawas, {\it unlike right sneutrinos}, and as a consequence neutrino physics is very relevant for them.
For example, if we work with 
Eq.~(\ref{evenLLL2}) 
using $T^{\nu}_{i} = A^{\nu}_{i} Y^{\nu}_{i}$, and
assuming the simplest situation that all the $A^{\nu}_i$ are naturally of the order of the TeV, neutrino physics determines sneutrino masses through the prefactor
$\frac{m_{{\mathcal{D}_i}}}{v_{iL}/\sqrt 2} = Y^{\nu}_i v_u/{v_{iL}}$.
Representative solutions for neutrino physics using diagonal neutrino Yukawas were summarized
in Subsec.~\ref{lightn} below Eq.~(\ref{neutrinoph2}).
Taking them into account, different hierarchies among the generations 
of left sneutrinos are possible 
using different hierarchies among $Y^{\nu}_i$ (and also $v_{iL}$).
According to the above discussion about the prefactor, and using the same classification as in Subsec.~\ref{lightn}, one obtains for the three
relevant cases:

1) $M<0$, with $Y^{\nu}_1<Y^{\nu}_2, Y^{\nu}_3$, and $v_{1L} > v_{2L}, v_{3L}$.

\noindent 
$m_{\widetilde{\nu}_{1L}}$ is the smallest of all the sneutrino masses.

2) $M>0$, with $Y^{\nu}_3 < Y^{\nu}_1 < Y^{\nu}_2$, and $v_{1L}<v_{2L}\sim v_{3L}$.

\noindent 
$m_{\widetilde{\nu}_{3L}}$ is the smallest of all the sneutrino masses.

3) $M>0$, with $Y^{\nu}_2 < Y^{\nu}_1 < Y^{\nu}_3$, and $v_{1L}<v_{2L}\sim v_{3L}$.

\noindent 
$m_{\widetilde{\nu}_{2L}}$ is the smallest of all the sneutrino masses.

{
Let us point out nevertheless concerning this qualitative analysis, that when off-diagonal neutrino Yukawas are allowed, it is not possible to arrive to a general conclusion regarding the hierarchy in {sneutrino} masses, specially when the gaugino seesaw is sub-dominant. This is because one can play with the hierarchies among $v_{iL}$ with enough freedom in the neutrino Yukawas in order to reproduce the 
experimental results. Therefore, there is no a priori knowledge of the hierarchies in 
the sneutrino masses, and carrying out an analysis case by case turns out to be
necessary. 
}

There is enough freedom in the parameter space of the $\mn$
in order to get {\bf heavy as well as light left sneutrinos} from Eq.~(\ref{evenLLL2}), and the latter scenario 
with the left sneutrino as the LSP was considered in Refs.~\cite{Ghosh:2017yeh,Lara:2018rwv,Kpatcha:2019gmq}. 
Thanks to RPV they decay to SM particles evading missing energy searches.
Due to the doublet nature of the left sneutrino, masses smaller than half of the mass of the SM-like Higgs were found to
be forbidden~\cite{Kpatcha:2019gmq} 
to avoid dominant decay of the latter into sneutrino pairs, leading to an inconsistency
with Higgs data.
In the analyses of Refs.~\cite{Lara:2018rwv,Kpatcha:2019gmq}, the
solutions of {type 2} were the ones interesting to be able to compare the signals with current LHC data. These issues will be discussed in more detail in Subsec.~\ref{leftsneutrinos}.






\subsubsection{Charged Scalars}
\label{chargeds}


Charged Higgses mix with left and right charged sleptons, thus the charged scalars
have the composition
$C^T=
({H^-_d}^*,{H^+_u},\widetilde{e}_{iL}^*,\widetilde{e}_{jR}^*)$, and
one obtains the mass terms, 
$-{C^*}^T 
{m}^2_{H^+} C$, with $m_{H^+}^2$ the $8\times 8$ (symmetric) `charged-Higgs' mass matrix.
Similar to the 
Higgs mass matrices where some sectors are decoupled, the $2\times 2$ 
charged Higgs submatrix is decoupled from the $6\times 6$ charged slepton 
submatrix. 
Thus, as in the MSSM, the mass of the charged Higgs 
is similar to the one of the doublet-like neutral pseudoscalar,
specially when the latter is not very mixed with the right sneutrinos. In this case, both masses are also similar to the
one of the heavy doublet-like neutral Higgs.

Concerning the $6\times 6$ submatrix, the right sleptons are decoupled from the left ones,
since the mixing terms are suppressed by $Y^{e}_{ij}$ or $v_{iL}$.
Then, the masses of right and left sleptons are basically determined by their corresponding soft terms, 
$m_{\widetilde{e}_{iR}}^2$ and
$m_{\widetilde{L}_{i}}^2$, respectively.
Although the left sleptons are in the same $SU(2)$ doublet as the
left sneutrinos, they are a little heavier than the latter mainly due to the
mass splitting produced by the D-term contribution, $-m_W^2 \cos 2\beta$.

\subsection{Charged Fermions}
\label{cfermions}




The MSSM charginos mix with the charged leptons in the $\mn$. 
In the basis 
$(\mathbf{\chi}^-)^T = ({(e_{iL})^c}^*, \widetilde{W}^-, \widetilde{H}^-_d)$  
and 
$(\mathbf{\chi}^+)^T = (e^*_{iR}, \widetilde{W}^+, \widetilde{H}^+_u)$, 
one obtains the mass terms, $-({\chi^{-}})^T {m}_{e}\chi^+ + \mathrm{h.c.}$, with
${m}_{e}$ the 
$5\times 5$ `lepton' mass matrix.
Nevertheless, the $2\times 2$ chargino submatrix is basically decoupled from the
$3\times 3$ lepton submatrix, since the off-diagonal entries are supressed 
by $Y^{\nu}_{ij}$, $Y^{e}_{ij}$, $v_{iL}$.
The former submatrix is like the one of the MSSM/NMSSM  
provided that one uses the effective $\mu$-term of Eq.~(\ref{eq:mu}).


\subsection{The Colored Sector}


\subsubsection{Quarks}

\noindent 
Down- and up-quark mass matrices are like the ones of the MSSM.



\subsubsection{Squarks}

Left and right down-squarks are mixed, thus they have the composition ${\tilde d}^{^T}=(\widetilde{d}_{iL},\widetilde{d}_{jR})$, and one obtains the mass terms,
$- {\tilde d}^{^T} 
{m}^2_{\widetilde d}\ {\widetilde d}^*$, with
${m}^2_{\widetilde d}$ the $6\times 6$ (symmetric) mass matrix. 
The same discussion applies to up-squarks.
These squark mass matrices,
when compared to the MSSM/NMSSM case, maintain their structure essentially unaffected, provided that one uses the effective $\mu$-term of Eq.~(\ref{eq:mu}), and neglects
the terms proportional to small parameters such as $Y^{\nu}_{ij}$, $v_{iL}$.


\subsubsection{Gluinos}

\noindent This sector has the same structure as in the MSSM.



\subsection{Electroweak Gauge Bosons}

The EW gauge sector has the same structure as in the SM. Note nevertheless that nonstandard
on-shell decays of $W^{\pm}$ and $Z$ bosons are possible within the framework of the $\mn$
compared to the MSSM~\cite{Ghosh:2014rha}. These modes are typically encountered in regions of the
parameter space with light singlet-like scalars, pseudoscalars, and neutralinos.
We will return to this scenario in Subsec.~\ref{unusual}.
Besides, in Subsec.~\ref{gauge} we will discuss the possibility of a new $Z'$ gauge boson in the context of extensions of the $\mn$ with an extra $U(1)'$ gauge group.





\section{Phenomenology}
\label{sec:searches}

The search for new physics is one of the most important experimental ventures in contemporary times. The $\mn$ provides a simple framework for this search, 
where the inclusion of RH neutrinos connects neutrino physics with RPV since the amount of the latter is determined by the neutrino Yukawas. Thus,
{\bf the smallness of neutrino masses is directly related with the low decay width of the LSP}.
Together with a very {\bf rich Higgs sector}, this gives a perfect combination for exciting new signals. 
It is true that,
besides the discovery of a particle compatible with the 
SM Higgs~\cite{Chatrchyan:2012xdj,Aad:2012tfa}, no hints for SUSY have been detected at the LHC yet, despite of numerous searches and tremendous efforts. 
But it is also important to realize that these searches for sparticles have assumed mainly signals with MET, inspired in RPC models such as the 
MSSM where the neutralino LSP is stable. This assumption is not possible in the $\mn$, where 
the
LSP
is not stable, making mandatory a reanalysis of the LHC data for this model.
%
In the next subsections, 
we will 
review the analyses of new signals in collider and non-collider experiments motivated by the model, that can be found in the literature so far.

\subsection{Searching for Sparticles at the LHC}
\label{sec:LHCsearchs}

We review in this subsection relevant signals at the LHC produced by several LSP candidates in the
$\mn$, that can be found in the literature.
The strategy employed to search for sparticles is to perform analyses of the parameter space associated to a particular LSP candidate, {\bf imposing compatibility with current experimental data on neutrino and Higgs physics as well as flavor with observables such as 
$B$ and $\mu$ decays}.
Let us emphasize that this strategy is crucial given the structure of the model, where neutrino and LSP physics are directly connected. In other words, we cannot assume a candidate for LSP with its corresponding new signals at the LHC without checking {at the same time whether} the neutrino data is reproduced. The two issues are not independent. 
This makes the $\mn$ more predictive, but also more complicated to analyze. 
{In the final subsections, we will also discuss unusual Higgs and EW gauge boson decays, which might help to probe the $\mn$ at the LHC.}

\subsubsection{Left Sneutrino} 
\label{leftsneutrinos}

We have explained in Subsec.~\ref{lhs} that
{left sneutrinos are special} in the $\mn$. This is because neutrino physics is very relevant for them,
determining their masses.
The first analysis of signals at the LHC of the left sneutrino LSP in this model was carried out in Ref.~\cite{Ghosh:2017yeh}, where the prospects for detection of signals
with diphoton plus leptons, diphoton plus MET from neutrinos, or multi-leptons, 
from the pair production of left sneutrinos/sleptons and their {\bf prompt decays} ($c\tau\lsim$ 0.1 mm), were analyzed.
A significant evidence is expected in the mass range of about 100 to 300 GeV, suggesting that
they deserve experimental attention. 
In the case of the multilepton signal, there exist generic searches for production of three or more leptons, which include also signal regions with a low missing transverse momentum and total transverse energy (see Refs.~\cite{Chatrchyan:2012mea,Chatrchyan:2014aea}). 
These searches are close to be sensitive to the sneutrino signal, and {\bf an updated analysis with current data could put constraints on the sneutrino LSP scenario}.

\begin{figure}[t!]
\centering
\label{fig:production31a} 
{\includegraphics[scale=0.27]{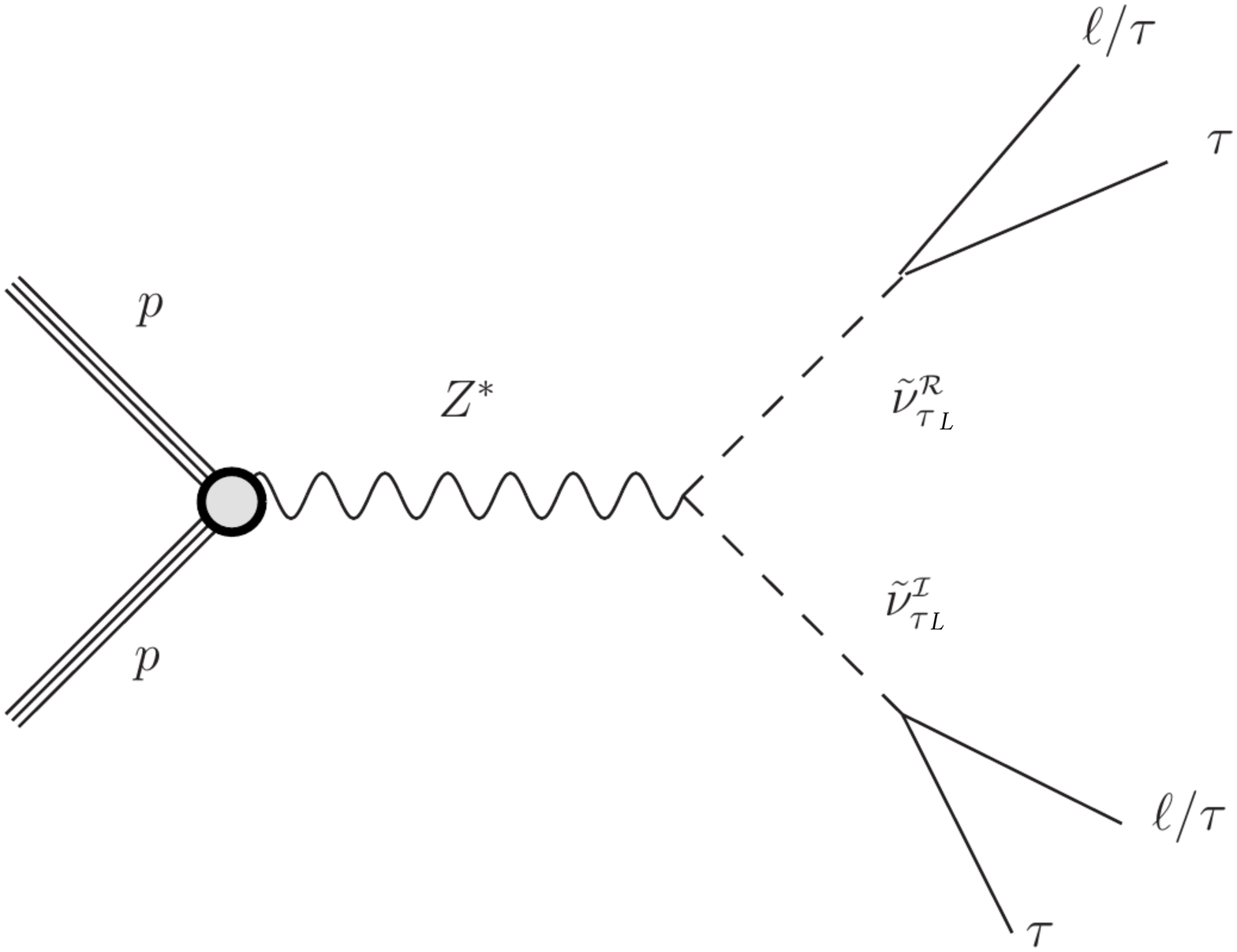}}\\ 
\vspace*{2mm}
{\includegraphics[scale=0.27]{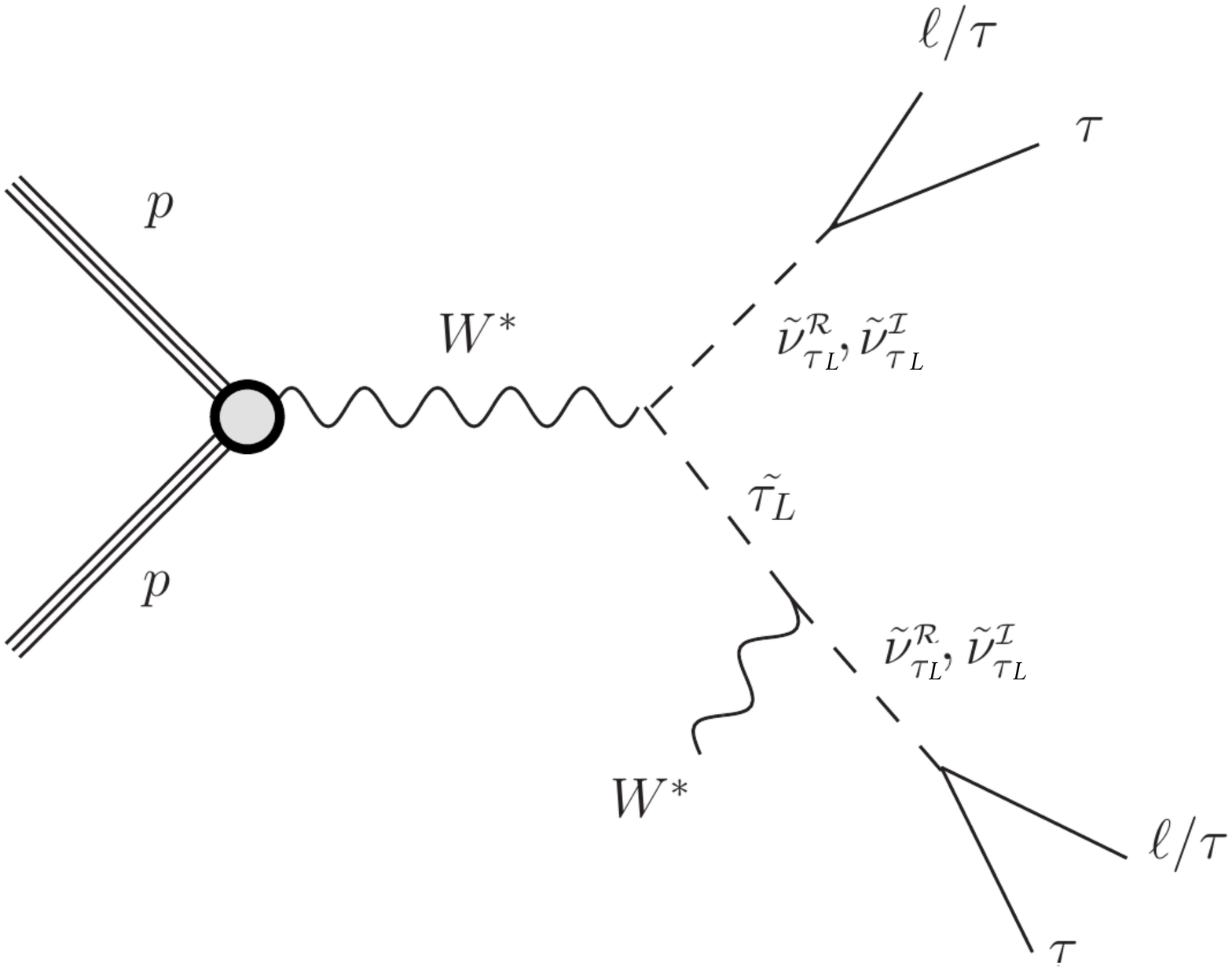}}\\
\vspace*{2mm}
{\includegraphics[scale=0.27]{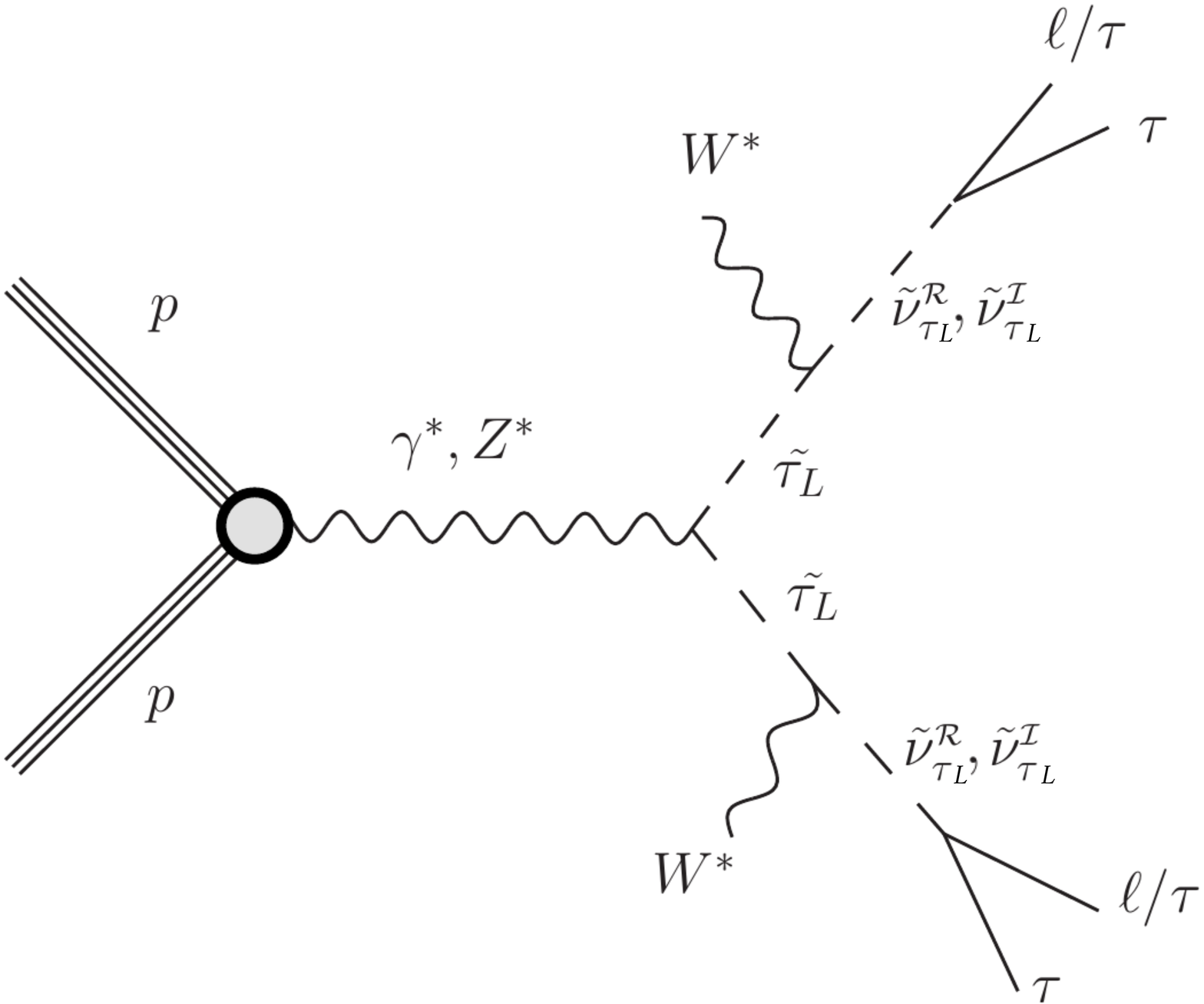}} 
\vspace*{2mm}
\caption{RPV decay channels into two $\tau\,\ell/\tau$, from a pair production at the LHC of scalar and pseudoscalar tau left sneutrinos co-LSPs via $Z$ channel, $\gamma$ and $Z$ channels, and $W$ channel, from
Refs.~\cite{Lara:2018rwv,Kpatcha:2019gmq}. Decay channels into one $\tau\,\ell/\tau$ plus neutrinos are the same but substituting 
one of the two vertices by a two-neutrino vertex.
The symbol $\ell$ is used for an electron or a muon, $\ell=e,\mu$,
and charge conjugation of fermions is to be understood where appropriate.
}
\label{fig:Tau-left-sneutrino-production.png}
\end{figure} 

The mass range of 45 to 100 GeV (with the lower limit imposed not to disturb the decay width of the $Z$) was covered in Refs.~\cite{Lara:2018rwv,Kpatcha:2019gmq}
 for the tau left sneutrino ($\widetilde{\nu}_{\tau L}$) LSP, with
{\bf displaced-vertex} searches.
In particular,
the decays of the $\widetilde{\nu}_{\tau L}$
LSP producing signals with dilepton pairs were studied,
using the {\bf present data set of the ATLAS 8-TeV search}~\cite{Aad:2015rba}
for long-lived particles using displaced lepton pairs $\ell\ell$ (with $\ell = e,\mu$),
as well as the prospects
for the 13-TeV searches.
The case of the $\widetilde{\nu}_{\tau L}$
LSP turns out to be particularly interesting because of the large value of the tau Yukawa coupling. A large fraction decays into a pair of tau leptons or a tau lepton and a light charged lepton, while the rest decays into a pair of neutrinos.
Thus, these processes can give rise to significant branching ratios (BRs) for 
decays to
$\tau\tau$ and 
$\tau \ell$,
once the sneutrinos are dominantly pair-produced via a Drell-Yan process mediated by a virtual
$W$, $Z$ or $\gamma$, as shown in Fig.~\ref{fig:Tau-left-sneutrino-production.png}.
Note that in this scenario the direct production of sleptons and their decays is a significant source of sneutrinos. This is because the
{\bf left stau} can be naturally the next-to-LSP (NLSP), since it is only a little heavier than the $\widetilde{\nu}_{\tau L}$ 
as discussed in Subsec.~\ref{chargeds}.
Thus, the stau has a dominant RPC prompt decay into 
a (scalar or pseudoscalar) sneutrino plus an off-shell $W$ producing a
soft meson or a pair of a charged lepton and a neutrino.

Subsequently, the pair-produced $\widetilde{\nu}_{\tau L}$ can decay
into $\tau\,\ell/\tau$.
As a result of the {\bf mixing between left sneutrinos and Higgses}, 
the sizable decay of $\widetilde{\nu}_{\tau L}$ into $\tau\tau$ is possible
because of the large value of the tau Yukawa coupling.  
Other sizable decays into $\tau\,\ell/\tau$ can occur through the Yukawa interaction of 
$\widetilde{\nu}_{\tau L}$ with $\tau$ and charged Higgsinos,
via the {\bf mixing between the charged Higgsinos and $\ell$ or $\tau$}.
To analyze these processes one can write approximate formulas for
the partial decay widths of the scalar/pseudoscalar tau left sneutrino.
The one
into 
$\tau\tau$ is given by: 
%
\begin{equation}
  \Gamma \left(\widetilde{\nu}_{\tau L}
\rightarrow 
 \tau\tau\right) 
\approx\frac{m_{\widetilde{\nu}_{\tau L}
}}{{16\pi}}
 \left(Y^\tau 
 Z^{H/A}_{\widetilde{\nu}_{\tau L} H_d}
- Y^{{\nu}_{\tau}} 
\frac{Y^{\tau}
}{3\lambda}
\right)^2,
  \label{eq:3.220}
\end{equation}
where $Y^{\tau}\equiv Y^e_{33}$, 
and $Z^{H/A}$ is the matrix which diagonalizes the mass matrix for the neutral 
scalars/pseudoscalars.  
The latter is
determined by the neutrino Yukawas, which are the order parameters of
the RPV. The contribution of $\lambda$ in the second term of
Eq.~(\ref{eq:3.220}) is due to the charged Higgsino mass that can be
approximated by the value of 
$\mu=3\lambda \frac{v_{R}}{\sqrt 2}$. 
In this computation, we are assuming for simplicity that $\lambda_i = \lambda$, $v_{iR}= v_{R}$, and
$\kappa_{iii}\equiv\kappa_{i}=\kappa$ and vanishing otherwise.
The partial decay width into $\tau
\ell$ can then be approximated for both sneutrino states by the second
term of Eq.~(\ref{eq:3.220}) with the substitution
$Y^{\nu_{\tau}}\rightarrow Y^{\nu_{\ell}}$: 
\begin{equation}
  \Gamma \left(\widetilde{\nu}_{\tau L}
\rightarrow 
\tau \ell\right) 
\approx\frac{m_{\widetilde{\nu}_{\tau L}
}}{{16\pi}}
\left(Y^{\nu_{\ell}}
\frac{Y^{\tau} 
}{3\lambda}
\right)^2.
  \label{eq:3.22}
\end{equation}

{On the other hand, 
the gauge interactions of 
$\widetilde{\nu}_{\tau}$ with neutrinos and binos (winos)
can produce a large decay width into neutrinos,
via the gauge {\bf mixing between gauginos and neutrinos}.}
This partial decay width
can be approximated for scalar and pseudoscalar sneutrinos as
\begin{eqnarray}
\sum_{i}{\Gamma(\widetilde \nu_{\tau L} \to \nu_\tau \nu_i  )} \approx 
\frac{m_{\widetilde \nu_{\tau L}}}{16\pi}
 \sum_{i}{\left|\frac{g'}{2}U^V_{i4}{-}\frac{g}{2}U^V_{i5}\right|^2},
 \label{--sneutrino-decay-width-2nus}
\end{eqnarray}
where $U^V$ is the matrix which diagonalizes the mass matrix for the 
neutral fermions, and the above entries can 
be approximated as 
\begin{eqnarray}
U^V_{i4}\approx\frac{{-}g'}{\sqrt{2}M_1}\sum_{l}{v_{lL}U^{PMNS}_{il}},\nonumber\\
U^V_{i5}\approx\frac{g}{\sqrt{2}M_2}\sum_{l}{v_{lL}U^{PMNS}_{il}}.
\label{--sneutrino-decay-width-2nus2}
\end{eqnarray}
Here $U^{PMNS}_{il}$ are the entries of the PMNS matrix, with
$i$ and $l$ neutrino physical and flavor indices, respectively. 
The relevant diagrams for  
$\widetilde{\nu}_{\tau L}$
searches that include this decay mode are the same as in
Fig.~\ref{fig:Tau-left-sneutrino-production.png}, but 
substituting one of the $\tau\,\ell/\tau$
vertices by a two-neutrino vertex.

There is enough freedom in the parameter space of the $\mn$ 
in order to get light left sneutrinos.
Assuming that the $A^{\nu}_i$ are 
naturally of the order of the TeV, values of the prefactor of
Eq.~(\ref{evenLLL2})
${Y^{\nu}_i v_u}/{v_{iL}}$ in the range of about $0.01-1$, i.e.
$Y^{\nu}_i\sim 10^{-8}-10^{-6}$, will give rise to left sneutrino masses in the
range of about $100-1000$ GeV.
Thus, with the hierarchy of neutrino Yukawas 
$Y^{\nu}_{3}\sim 10^{-8}-10^{-7}<Y^{\nu}_{1,2}\sim 10^{-6}$, we can obtain a
 $\widetilde{\nu}_{\tau L}$ LSP with a mass around 100 GeV whereas the masses of
 $\widetilde{\nu}_{e,\mu}$ are of the order of the TeV.
Clearly, we are in the case of solutions for neutrino physics of {type 2} discussed in Subsec.~\ref{lhs}. 
Actually this type of hierarchy, with significant values for $Y^{\nu}_{1,2}$,
increases the dilepton BRs of the $\widetilde{\nu}_{\tau L}$ LSP producing signals that can be probed at the LHC.
The most important contribution to BRs comes
from the channel $\widetilde{\nu}_{\tau L}\rightarrow \tau\mu$.
It is found then that
the decay distance of the left sneutrino tends to be {\bf as large as $\gtrsim
1$~mm}, which thus can be a good target of displaced vertex searches. 

The strategy employed to search for these points was {\bf to perform scans} of the parameter space of this scenario imposing compatibility with current experimental data on neutrino and Higgs physics 
(using 
{\tt HiggsBounds}~\cite{Bechtle:2008jh,Bechtle:2011sb,Bechtle:2013gu,Bechtle:2013wla,Bechtle:2015pma} and 
{\tt HiggsSignals}~\cite{Bechtle:2013xfa,Stal:2013hwa,Bechtle:2014ewa}), as well as with flavor observables such as $B$ and $\mu$ decays. To carry it out, 
{\bf a powerful likelihood data-driven method} based on the algorithm called {\tt Multinest}~\cite{Feroz:2007kg,Feroz:2008xx,Feroz:2013hea} was used~\cite{Kpatcha:2019gmq}.
To compute the spectrum and the observables, a suitable modified version of the {\tt SARAH} code~\cite{Staub:2013tta} to generate a 
{\tt SPheno}~\cite{Porod:2003um,Porod:2011nf} version for the model was used.

In order to efficiently scan the model, 
it is important to identify first
the parameters to be used, and optimize their number and their ranges of values.
The relevant independent parameters in the neutrino sector of the $\mn$ are
$\lambda, \kappa, v_R, v_{iL}, Y^{\nu}_i, \tan\beta$, $M_1$ and $M_2$ (see Eq.~(\ref{freeparametersn})).
Since $\lambda, \kappa$ and  $v_R$ are crucial for Higgs physics 
(see Eq.~(\ref{freeparameters})), one can fix first them to appropriate values. The parameter
$\tan\beta$ 
is also important for both, Higgs and neutrino physics, thus one can consider a narrow range of possible values to ensure good Higgs physics.
Inspired by GUTs one can assume $M_2 = 2M_1$, scan over $M_2$ and use
$M$, which is a kind of average of bino and wino soft masses (see Eq.~(\ref{effectivegauginomass2})), 
as the relevant parameter.
On the other hand, 
sneutrino masses introduce in addition the parameters 
$T^{\nu}_i$ (see Eq.~(\ref{evenLLL2})). In particular, $T^{\nu}_3$ is the most relevant one for the discussion of the $\widetilde{\nu}_{\tau L}$ LSP, and it is scanned in an appropriate range of small values. Since the left sneutrinos of the first two generations must be heavier, $-T^{\nu}_{1,2}$ are fixed to a larger value, $10^{-3}$.

\begin{table}
\begin{center}
\begin{tabular}{|l|l|}
\hline
 \multicolumn{1}{|l|}{\bf Scan 1 ($S_1$)}&\multicolumn{1}{l|}{ \bf Scan 2 ($S_2$)}\\
\cline{1-2}
  $\tan\beta \in (10, 16)$ & $\tan\beta \in (1, 4)$\\ 
\hline
\multicolumn{2}{|l|}{ \quad \quad $Y^{\nu}_{i} \in (10^{-8} , 10^{-6})$ }\\
\multicolumn{2}{|l|}{ \quad \quad $v_{iL} \in (10^{-6} , 10^{-3})$  }\\
\multicolumn{2}{|l|}{ \quad \quad  $-T^{\nu}_{3} \in (10^{-6} , 10^{-4})$ }\\
\multicolumn{2}{|l|}{ \quad \quad $M_2 \in (150 , 2000)$ }\\
\cline{1-2}
\end{tabular}
\end{center}
  \caption{Range of low-energy values of the input parameters that are varied in the two scans, where  $Y^{\nu}_{i}$, $v_{iL}$, $T^{\nu}_{3}$ and $M_2$ are $\log$ priors (in logarithmic scale) while $\tan\beta$ is a flat prior (in linear scale), from Ref.~\cite{Kpatcha:2019gmq}.
The
VEVs $v_{iL}$, and the soft parameters $T^{\nu}_{3}$ and $M_2$, are given in GeV.}
 \label{Scans-priors-parameters}
\end{table}

Summarizing, scans over the 9 parameters
\bea
Y^{\nu}_i,\,\, 
v_{iL},\,\, 
T^{\nu}_3,\,\, 
\tan\beta,\,\, 
M_2,
\label{finalpp}
\eea
were performed in Ref.~\cite{Kpatcha:2019gmq} as shown in Table~\ref{Scans-priors-parameters}.
The ranges of $v_{iL}$ and $Y^{\nu}_i$ are natural in the context of the EW seesaw of the $\mn$.
The range
of $T^{\nu}_{3}$ is also natural if we follow the usual assumption based on the supergravity framework discussed in Eq.~(\ref{tmunu}) that the trilinear parameters are proportional to the corresponding Yukawa couplings, i.e. in this case $T^{\nu}_{3}= A^{\nu}_{3} Y^{\nu}_3$ implying 
$-A^{\nu}_{3}\in$ ($1, 10^{4}$) GeV.
Concerning $M_2$, its range of values is taken such that a bino at the bottom
of the neutralino spectrum leaves room to accommodate a 
$\widetilde{\nu}_{\tau L}$ LSP with a mass below 100 GeV. 
Scans 1 ($S_1$) and 2 ($S_2$) correspond  
to different values of $\tan\beta$, and other benchmark
parameters as shown in 
Table 3 of Ref.~\cite{Kpatcha:2019gmq}.
For example, two values of $\lambda$ are chosen in order to cover a representative region,
from
a small/moderate value, $\lambda\approx 0.1$ ($S_1$), to a large value, $\lambda\approx 0.4$  ($S_2$), in the border of perturbativity up to the GUT scale~\cite{Escudero:2008jg}.
With the help of {\tt Vevacious}~\cite{Camargo-Molina:2013qva}, it was also checked that the EWSB vacua corresponding to the previous allowed points are stable.

Due to the doublet nature of the left sneutrino, {\bf $m_{\widetilde \nu_{\tau L}}$ is constrained to be heavier than about 61 GeV},
which corresponds to half of the mass of the SM-like Higgs (allowing a $\pm$3 GeV theoretical uncertainty on its mass).
For smaller masses, the latter would dominantly decay into sneutrino pairs, {leading to an inconsistency with Higgs data}. {In this scenario 
the SM-like Higgs 
decays into pairs of scalar/pseudoscalar tau left sneutrinos via gauge interactions, 
mostly from D-terms $\sim \frac{1}{4}(g^2+g'^2)\widetilde \nu_i \widetilde \nu_i^*  H^0_uH^{0*}_u$,
since its largest component is $H^0_u$}.

\begin{figure}[t!]
 \centering
\includegraphics[width=\linewidth, height=0.35\textheight]{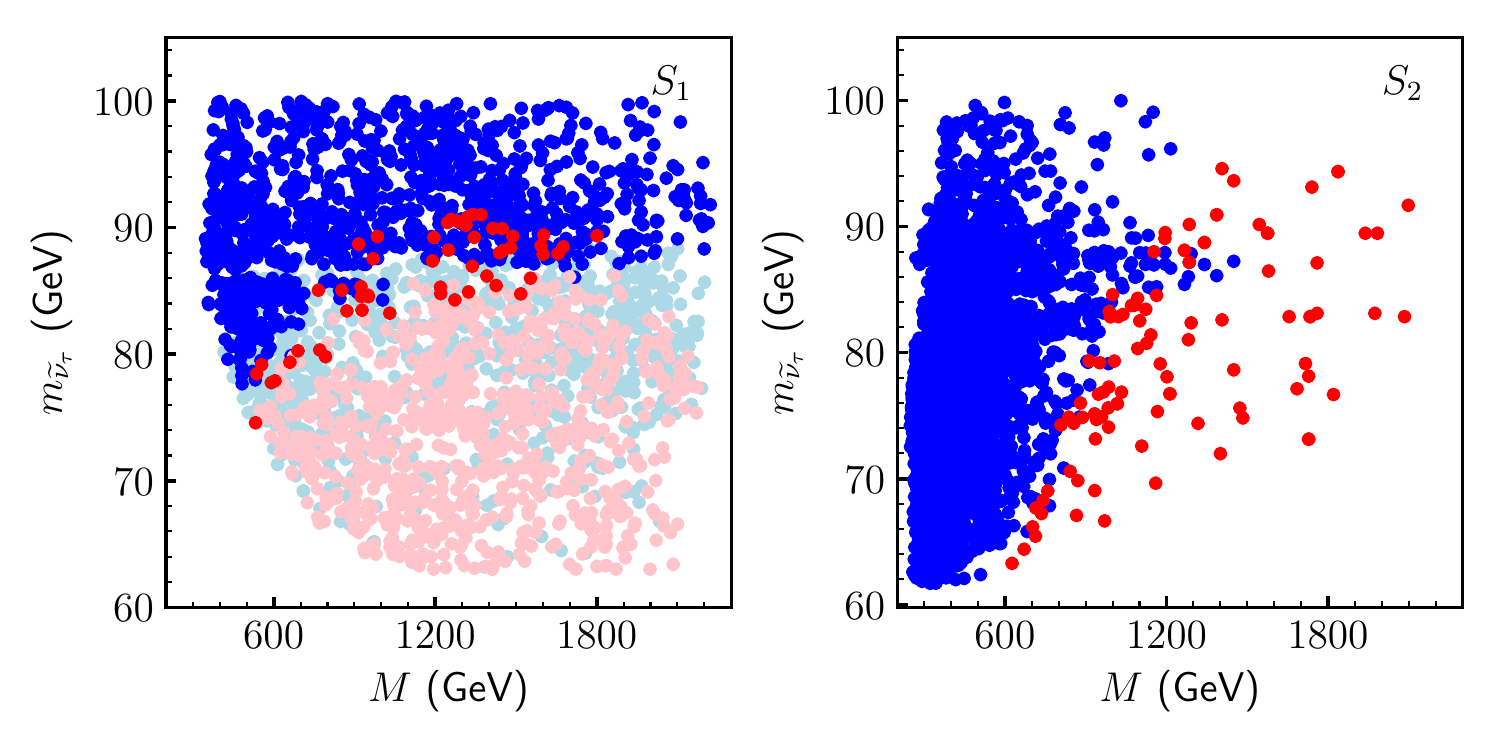}
 \caption{Tau left sneutrino LSP mass vs. the average gaugino mass $M$ (see Eq.~(\ref{gauginom})) 
 for the two relevant scans discussed in the text, $S_1$ (left) and $S_2$ (right),
from Ref.~\cite{Kpatcha:2019gmq}.
They cover a representative region:
from
a small/moderate value, $\lambda\approx 0.1$ ($S_1$), to a large value, $\lambda\approx 0.4$  ($S_2$), in the border of perturbativity up to the GUT scale.
In both plots,
 the dark-red points indicate that the number of signal events is
 above 3 and therefore detectable as discussed in the text, analyzing the prospects for the 13-TeV search with 
 an integrated luminosity of 300 fb$^{-1}$, combining
 the $\mu\mu$, $e\mu$ and $ee$ channels, and considering also the
 optimization of the trigger requirements discussed in Ref.~\cite{Lara:2018rwv}. The light-red points in scan $S_1$ although have a number of signal events above 3, are already excluded by the LEP result, as discussed in the text.
 The dark-blue points indicate that the number of signal events is below 3 and therefore inaccessible. The light-blue points in scan $S_1$ have also a number of signal events below 3, and, in addition, are already excluded by the LEP result.
}
 \label{Cut-MSvL3-M-Av3.png}
\end{figure}

To analyze the points fulfilling the above mass bound, one has to take into account that the search of the ATLAS collaboration for dilepton displaced vertices is basically background free, and they found no event~\cite{Aad:2015rba}.
However,
one of the {\bf problems with the existing searches} is that they are designed for a generic purpose and
therefore {are} not optimized for light metastable particles such as 
the $\widetilde{\nu}_{\tau L}$.
One cannot directly apply the limits provided by the ATLAS
collaboration 
to the left sneutrino
case, since the ATLAS analysis simulates the decay of a heavy gluino
into a light and a heavy neutralino. The former case represents a highly
boosted light particle decaying into a pair of muons, while the latter
represents a heavy non-boosted particle decaying in the same way. Yet,
the sneutrino features a light non-boosted particle. This analysis can be
extended, nevertheless, combining information from both situations
for {\bf recasting the ATLAS search} to the case of the $\widetilde{\nu}_{\tau L}$.

The final result of the analysis for the 8-TeV case~\cite{Kpatcha:2019gmq} is that no points of the parameter space of the $\mn$ can be probed, because all of them have a number of signal events below 3, which in this case is the number compatible at
the $2\sigma$ level with zero observed events.
This is also true even considering the optimization of the
trigger requirements proposed in Ref.~\cite{Lara:2018rwv} for which the number of background events
can still be regarded as zero.
Nevertheless, important regions can be probed at the LHC run 3 with the trigger optimization, as summarized in Fig.~\ref{Cut-MSvL3-M-Av3.png}.
There, the dark-red points indicate that the number of signal events is above 3, and therefore detectable, whereas the dark-blue points correspond to regions where the number of signal events is below 3.

{Note that the light-red points in scan $S_1$, as well as the light-blue points on top of the dark-blue ones, are already excluded by the {\bf LEP bounds} on left sneutrino masses~\cite{Abreu:1999qz,Abreu:2000pi,Achard:2001ek,Heister:2002jc,Abbiendi:2003rn,Abdallah:2003xc}. To carry out this analysis, one can consider e.g. Fig.~6a of Ref.~\cite{Heister:2002jc}, where the cross section upper limit for tau sneutrinos decaying directly to $\ell\ell\tau\tau$ via a dominant RPV operator, $\hat L\hat L\hat e^c$, is shown.
Assuming ${\rm BR} = 1$, a lower bound on the sneutrino mass of about 90 GeV was obtained through the comparison with the MSSM cross section for pair production of tau sneutrinos. 
To recast this result to the $\mn$ scenario, one multiplies this cross section by the factor
$\text{BR}(\widetilde{\nu}_{\tau L}^\mathcal{R}\to\tau\mu)\times
\text{BR}(\widetilde{\nu}_{\tau L}^\mathcal{I}\to\tau\mu)$ for each of the points studied.
For an average value of $\text{BR}(\widetilde{\nu}_{\tau L}\to\mu\mu)=0.1$ as one obtains in the scan of the
$\mn$, the cross section must be multiplied then by a factor of $\sim 0.33$, {\bf lowering the bound} on the sneutrino mass to {\bf about 74 GeV} in the $\mn$ scenario, as can be seen in Fig.~\ref{Cut-MSvL3-M-Av3.png}.

Concerning scan $S_2$, the BRs into charged leptons are about two orders of magnitude smaller than for $S_1$, and therefore following the above discussion we have checked that 
{\bf no points are excluded by LEP} results in this case.
Note that although these BRs are smaller, still a significant number of points with signal events above 3 can be obtained when $M$ increases because of the larger value of the decay length, which gives rise to a larger vertex-level efficiency.

\subsubsection{Charged Slepton} 

There are many LEP searches for staus~\cite{Abreu:1999qz,Abreu:2000pi,Achard:2001ek,Heister:2002jc,Abbiendi:2003rn,Abdallah:2003xc}. 
For example, we have seen in the previous subsection that the {left stau} does not decay directly but through an off-shell $W$ and a $\widetilde{\nu}_{\tau L}$. Therefore, searches for the direct decay of left sleptons are not relevant in the $\mn$.
It is then convenient to study them together with the
left sneutrino LSP, as was carried out in Refs.~\cite{Ghosh:2017yeh,Lara:2018rwv,Kpatcha:2019gmq} and discussed in the previous subsection. 

Concerning a right slepton LSP, there are no analyses of this scenario in the $\mn$ yet.

\subsubsection{$\mn$ Neutralino} 
\label{neutralino}

We discussed in Subsec.~\ref{neutralinos} that 
the lightest neutral fermions are the three light LH neutrinos. 
The rest of the (heavy) fermions are the usual neutralinos of the MSSM mixed with the RH neutrinos. We denoted these seven states as $\mn$ neutralinos.
Decays of $\mn$ neutralino LSP to two-body ($W\ell$, $Z\nu$) and 
three-body 
($\ell q_i \bar q_j$, $\ell_i \ell_j\nu$, $\nu q \bar q$, 3$\nu$)
final states were 
calculated in Refs.~\cite{Ghosh:2008yh,Bartl:2009an}.
As an example, we show in Fig.~\ref{Imagen1} from Ref.~\cite{Bartl:2009an} the average decay length in meters of the lightest $\mn$ neutralino as a function of the mass for different scenarios.
For neutralino masses below about 50 GeV decay lengths become {\bf larger than 1 meter}, implying that 
a large fraction of neutralinos will decay outside typical collider detectors. Nevertheless,
if one allows for lighter scalars/pseudoscalars in such a way that the neutralino can decay to 
scalar/pseudoscalar plus neutrino, the average decay length can be easily reduced by several
orders of magnitude.
Characteristics correlations of the decay BRs with the measured neutrino angles were also analyzed in those works.

\begin{figure}[t!]
 \begin{center}
       \epsfig{file=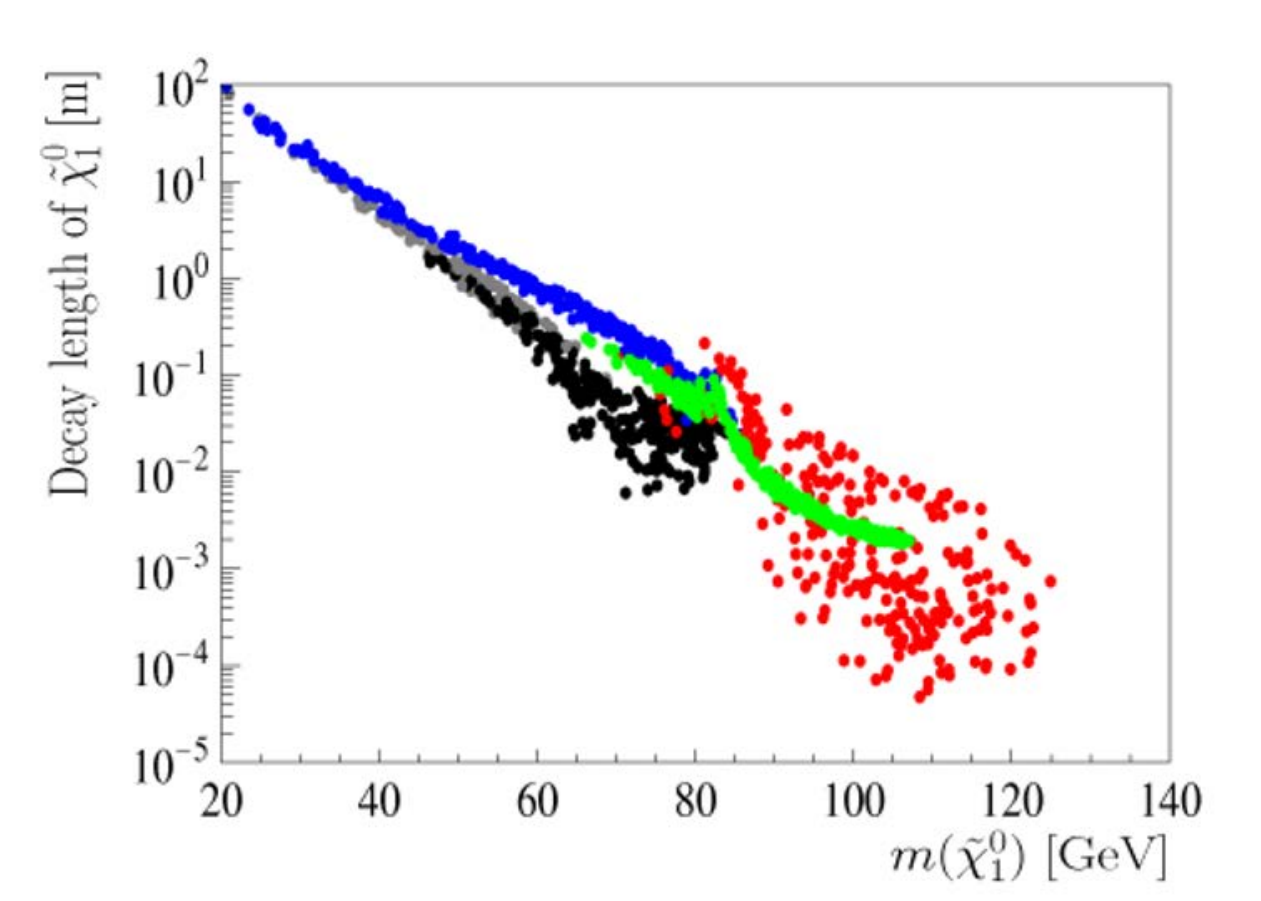,height=6.0cm}
\captions{
Decay length of the $\mn$ lightest neutralino $\tilde\chi^0_1$ in m as a function of its mass 
$m(\tilde\chi^0_1)$ in GeV for different values of the parameters of the model, chosen in such a way that no
lighter scalar/pseudoscalar states with masses smaller than $m(\tilde\chi^0_1)$ appear,
from
Ref.~\cite{Bartl:2009an}.
Different colors stand for benchmark points (see Ref.~\cite{Bartl:2009an} for details) with real singlino $|N_{45}|^2>0.5$ (gray), mixture state (black),
real singlino (blue), mixture state (red), mixture state (green).}
    \label{Imagen1}
 \end{center}
\end{figure}

In Refs.~\cite{Fidalgo:2011ky,Ghosh:2012pq,Ghosh:2014ida}, the signatures produced 
through two-body Higgs decays into $\mn$ neutralinos were analyzed in detail.
Special attention was paid to decays of the neutralino via a light scalar/pseudoscalar.
Depending on the associated decay length, the distinctive signal can be displaced multi-leptons/jets/photons with small/moderate MET from neutrinos.
{Signatures through unusual $W^{\pm}$ and $Z$ decays were also studied in
Ref.~\cite{Ghosh:2014rha}, and we will discuss this issue in more detail in 
Subsec.~\ref{unusual}.}

  \begin{figure}[]
 \centering
 \includegraphics[width=218pt,keepaspectratio=true]{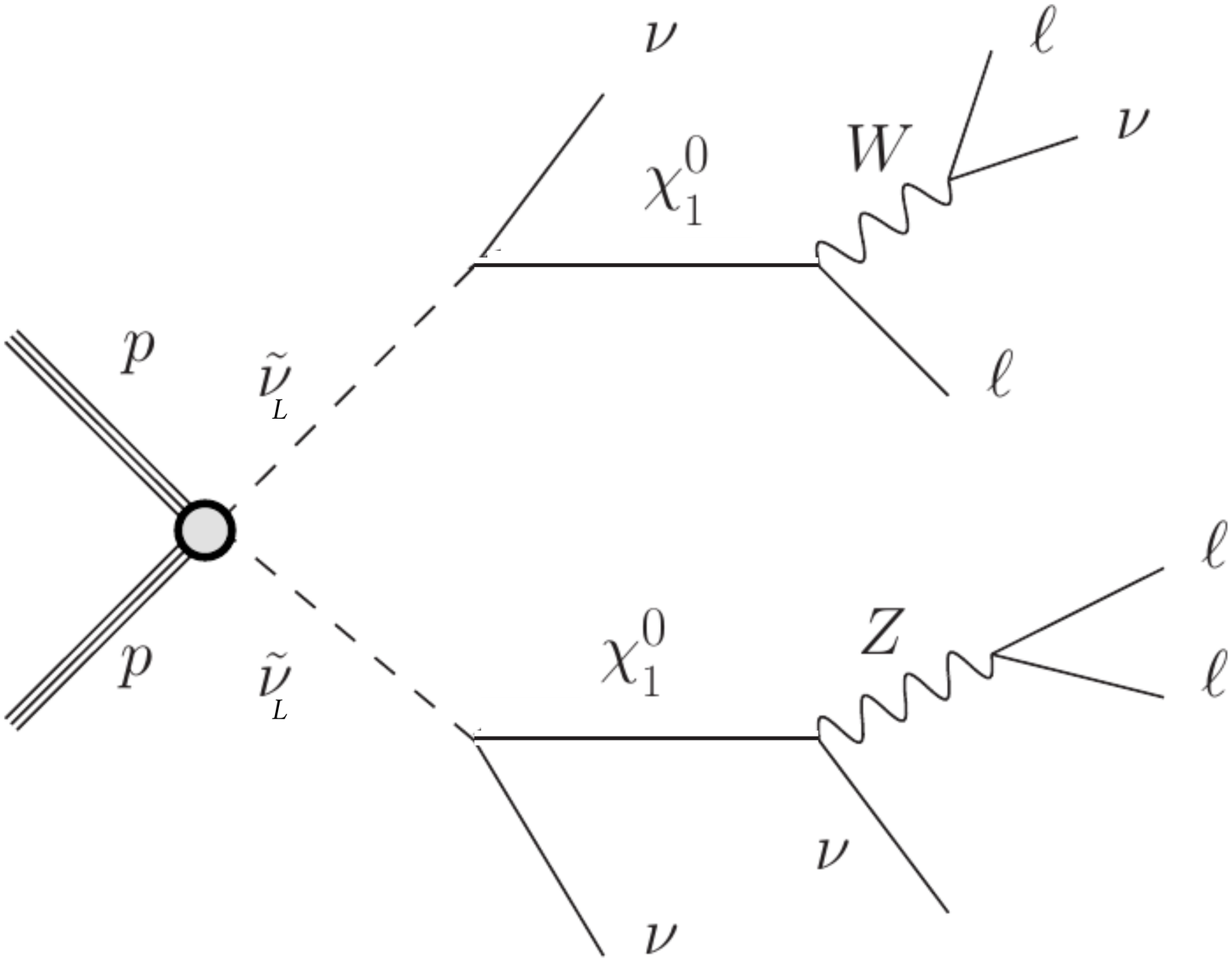}\\\vspace{20pt}
 \includegraphics[width=218pt,keepaspectratio=true]{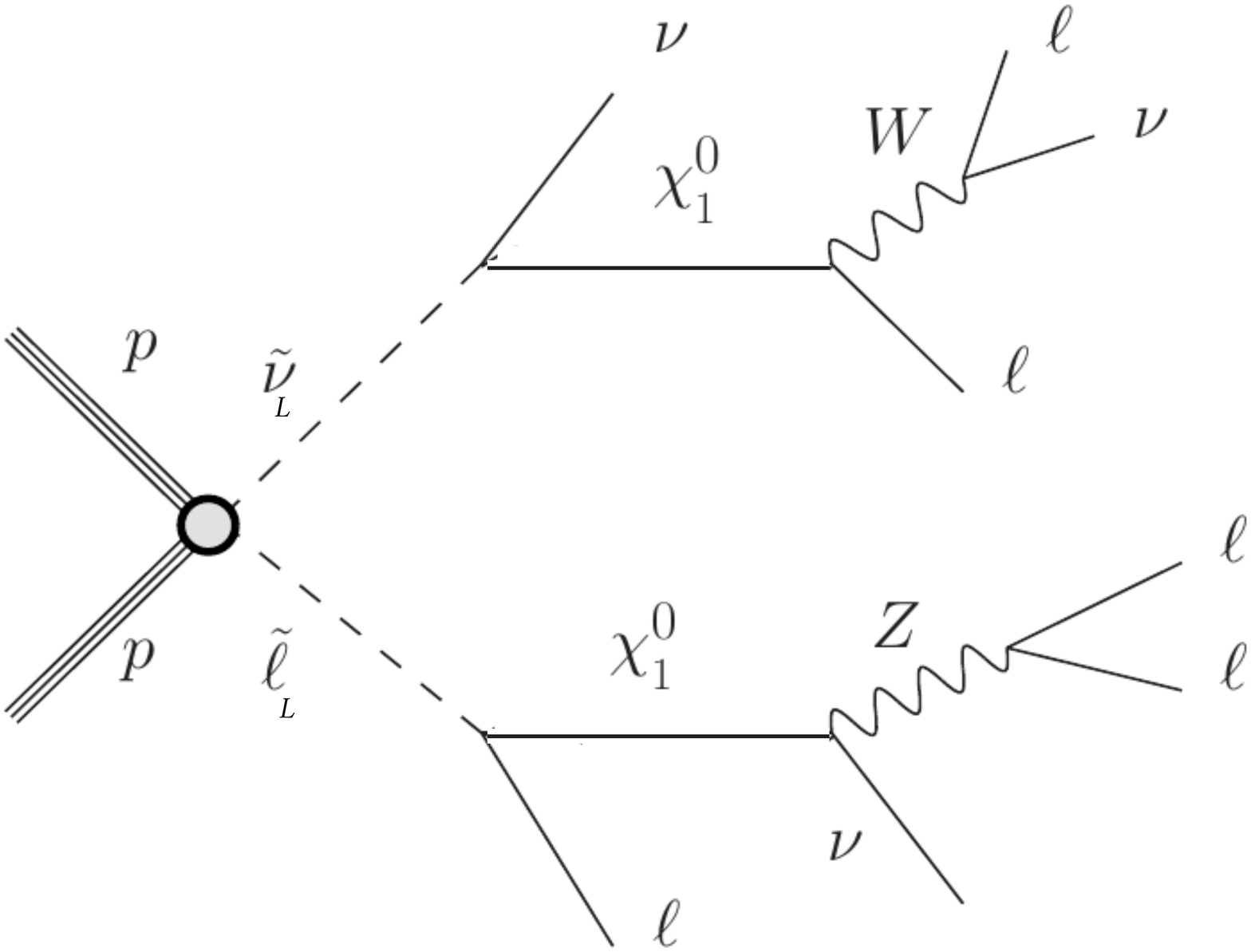}\\\vspace{20pt} 
\includegraphics[width=218pt,keepaspectratio=true]{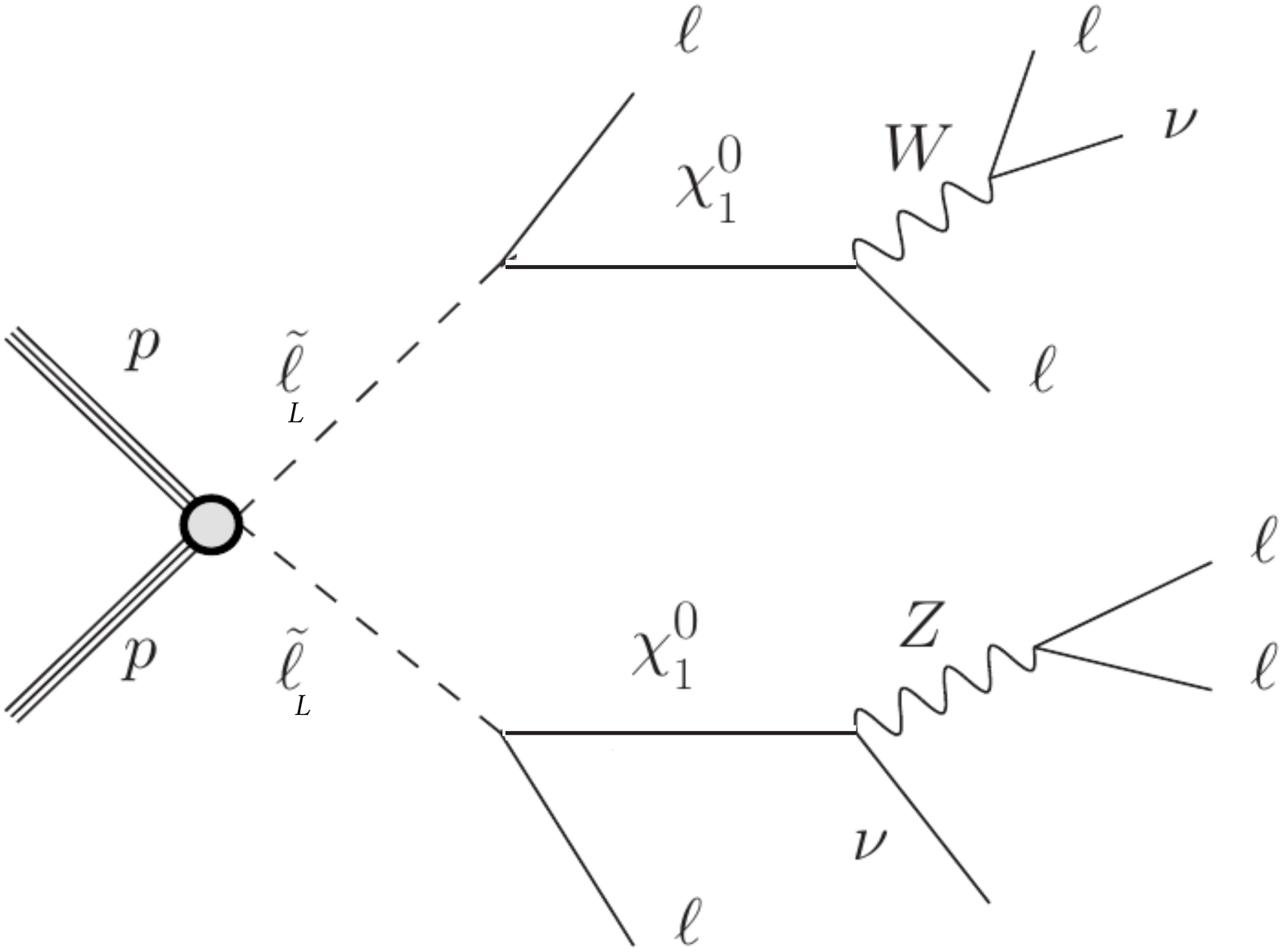}\\\vspace{20pt}
 \caption{Relevant diagrams of the benchmark $\mn$ scenario  
of RPC left sneutrino/charged slepton pair production, followed by the RPV decay of the 
bino-like LSP, $\tilde{\chi}^0_1$,
from Ref.~\cite{Lara:2018zvf}.
The symbol $\ell$ is used for electron, muon or tau
$\ell=e,\mu,\tau$,
and charge conjugation of fermions is to be understood where 
appropriate.
}
 \label{fig:production}
\end{figure}

The only comparison that has been made so far of a
{\bf $\mn$ neutralino with LHC data}, was carried out
in Ref.~\cite{Lara:2018zvf} using a bino-like LSP.
The pair production cross section of bino-like neutralinos at large hadron colliders is very small, since 
there is no direct coupling between the bino flavor state and the gauge bosons, and we are assuming that the rest of the spectrum remains decoupled. Binos are produced mainly through virtual $Z$ bosons in the s channel exploiting their small Higgsino flavor composition, or through the t channel interchange of virtual first generation squarks, strongly suppressed by their large masses.
Nevertheless, the bino-like LSP can be produced in the decay of other SUSY particles, which although heavier, have a higher production cross section at the LHC. That is the case when the left sneutrino is the NLSP. After production, the left sneutrinos decay to the bino LSP.
The dominant pair production channels of sleptons at hadron colliders have been already discussed in Subsec.~\ref{leftsneutrinos} (see also Fig.~\ref{fig:Tau-left-sneutrino-production.png}).
Then, the RPC decays ${\tilde{\nu}}_L\to\nu\tilde{\chi}^0_1$ and ${\tilde{\ell}}_L\to\ell\tilde{\chi}^0_1$ dominate 
over the RPV ones which are suppressed by the smallness of $Y^{\nu}$,
  thereby pair production of sneutrinos/sleptons at the LHC will be a source of bino pairs.
  Note that although the left sneutrino is lighter than its corresponding left slepton as discussed in previous subsections, since the mass separation is always smaller than $m_W$, the phase space suppression makes the decay ${\tilde{\ell}}_L\to \ell\tilde{\chi}^0_1$ dominant. 
  
Subsequently, {\bf binos will decay 
mediated through the RPV mixing between the bino and neutrinos}
to $W\ell$ or $Z\nu$.
Approximate formulas for the partial decay widths are as follows:
\begin{equation}
\Gamma ({\tilde{\chi}^0\to W \ell_i})\approx\frac{g^2 m_{\tilde{\chi}^0}}{16\pi}\left(1-\frac{m^2_{W}}{m_{\tilde{\chi}^0}^2}\right)^2\left(1+\frac{m_{\tilde{\chi}^0}^2}{2m_W^2}\right)\left|U^{V}_{\tilde B\nu_i}\right|^2,
 \label{decay_wl}
\end{equation}
%
\begin{equation}
\sum_i \Gamma ({\tilde{\chi}^0\to Z \nu_i})\approx
\frac{g^2 m_{\tilde{\chi}^0}}{16\pi\cos^2\theta_W}\left(1-\frac{m^2_{Z}}{m_{\tilde{\chi}^0}^2}\right)^2\left(1+\frac{m_{\tilde{\chi}^0}^2}{2m_Z^2}\right)
\sum_i \left|\sum_j U^{V}_{\tilde B\nu_j}{U_{\nu_i\nu_j}^{V^*}}\right|^2,
 \label{decay_Znu}
\end{equation}
where $U^V$ is the matrix that diagonalizes the mass matrix for the neutral fermions~\cite{Escudero:2008jg,Ghosh:2017yeh}, and $U^{V}_{\tilde B\nu_i}$ can be approximated as 
$\frac{g' v_i}{M_1}$.
The relevant signals with multi-leptons (up to six charged leptons) plus MET from neutrinos, are shown in 
Fig.~\ref{fig:production}.
Given the small value of $M_1$ analyzed, in the range between $m_Z$ and 200 GeV, close to the masses of the gauge bosons, and the sneutrino VEVs used while producing an acceptable mass scale for the neutrinos,
one obtains an upper bound for the total width of the bino
$\gtrsim 10^{-12}$ GeV corresponding to a proper {\bf decay length $c\tau\lsim 0.2$ mm}. This decay length is short enough as to be able to compare this process with {\bf prompt ATLAS searches}~\cite{Aaboud:2018sua,Aad:2019vvi} for EW sparticles in RPC models assuming wino-like chargino-neutralino production with a bino-like LSP.

After {\bf recasting the ATLAS result} to the case of the sneutrino-bino scenario, 
{\bf only a small region of the parameter space of the $\mn$ 
was excluded}~\cite{Lara:2018zvf}. This is 
the region of bino (sneutrino) masses $110-150$ ($110-160$) GeV, as shown in
Fig.~\ref{fig:exclusions}.
There, we can also see the prospects for the 300 fb$^{-1}$ searches.


\begin{figure}[]
 \centering
 \includegraphics[scale=0.45,keepaspectratio=true]{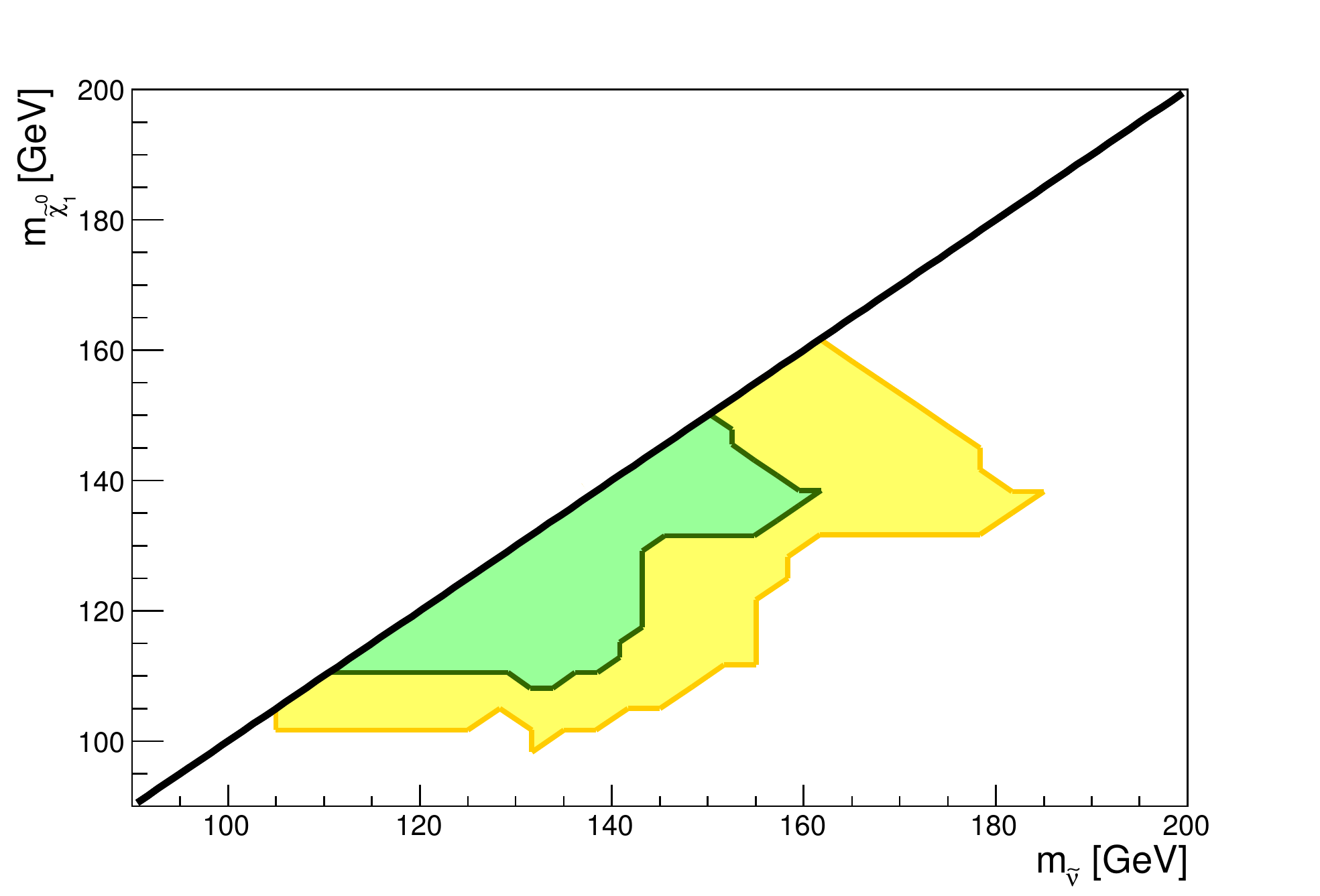}
\caption{Regions of the $\mn$ that will be probed by the signal with
three light leptons plus MET discussed in the text in the parameter space 
$m_{\tilde\nu}-m_{\tilde{\chi}^0_1}$ with universal sneutrino masses, for the 13-TeV search with an integrated luminosity 300 fb$^{-1}$ (yellow),
from Ref.~\cite{Lara:2018zvf}.
The green region is already excluded by 
the 139 fb$^{-1}$ data.
}
 \label{fig:exclusions}
\end{figure}

\subsubsection{Chargino}
 
We explained in Subsec.~\ref{cfermions} that although the MSSM charginos mix with
the charged leptons in the $\mn$, they are basically decoupled. 
The main difference being that in the MSSM the lightest chargino cannot be the LSP because it is a charged particle that would contribute to DM. 
This is not a problem in the $\mn$, where the chargino LSP is not stable. Although
this parameter space with a chargino lighter than neutralinos is not easy to be obtained, it deserves further attention. 

\subsubsection{Squark}
 
Let us remark that the MSSM limits on squark masses cannot be applied to the $\mn$. For example, if the stop is the LSP it can decay only via RPV channels into top plus neutrino and bottom plus lepton, and these decays can be prompt or displaced depending on the region of the parameter space of the model. Thus, dedicated analyses are necessary for recasting ATLAS and CMS results to the many possible cases of the model. This is work in progress~\cite{lara:2020xx}.


%
%
%
%

\subsubsection{Higgs decays}
\label{higgsd}

We already mentioned in Subsec.~\ref{neutralino} the signatures produced
through two-body Higgs decays into $\mn$ neutralinos studied in Refs.~\cite{Fidalgo:2011ky,Ghosh:2012pq,Ghosh:2014ida}.
In Refs.~\cite{Fidalgo:2011ky,Ghosh:2014ida}, a Higgs-like scalar decaying into a pair of light scalars/pseudoscalars producing final states with prompt multi-leptons/taus/jets/photons at the LHC, even via Higgs-to-Higgs cascade decays, was analyzed. 
It was also suggested in Ref.~\cite{Fidalgo:2011ky}, before the {discovery} of the Higgs-like scalar, the re-analysis of the LEP data in the light of the excess reported at a mass $\sim$ 96 GeV.
After the {discovery} of the Higgs boson at the LHC, the study of interesting benchmark points (BPs) in the $\mn$ with singlet-like eigenstates lighter than the SM-like Higgs boson was
carried out in Refs.~\cite{Biekotter:2017xmf,Biekotter:2019gtq}.
In addition, a simultaneous explanation of the two excesses measured at LEP and CMS at a mass 
$\sim$ 96 GeV, was also given.


In a recent work~\cite{Kpatcha:2019qsz}, the parameter space of the Higgs sector of the $\mn$ was scanned using the same strategy as in Ref.~\cite{Kpatcha:2019gmq}.
In particular, 
imposing compatibility with current experimental data on neutrino and Higgs physics 
(using 
{\tt HiggsBounds} and 
{\tt HiggsSignals}), as well as with flavor observables such as $B$ and $\mu$ 
{decays},
and sampling the model using a likelihood data-driven method based on 
{\tt Multinest}.
The result of the analysis was 
that the parameter space of the $\mn$ contains many viable solutions, including also many different phenomenological possibilities. Interesting BPs can be found in Appendix C
of Ref~\cite{Kpatcha:2019qsz}.
For example, there are solutions where the SM-like Higgs is the lightest scalar (see e.g. red and light-red points in Fig.~\ref{S1Lambda-kappa}), but also solutions where right sneutrino-like states are lighter (blue and light-blue points). In the latter case, it is even possible to have these (singlet-like) scalars with masses $\lsim m_h/2$.
In addition, there are also
solutions where several scalars are degenerated with masses close to 125 GeV, and can have their signals rates superimposed contributing to the resonance observed at 125 GeV.

Given these results, it is then important to study in detail the collider phenomenology of the solutions found. In particular, the impact of the new states, not only the right sneutrinos but also the left sneutrinos, and the neutralinos containing {RH} neutrinos. Novel signals associated to them might help to probe the $\mn$ at the LHC. 

%
%
%
%

\subsubsection{Unusual $W^{\pm}$ and $Z$ decays}  
\label{unusual}

In the framework of the $\mn$, nonstandard
on-shell decays of $W^{\pm}$ and $Z$ bosons are possible~\cite{Ghosh:2014rha}.
These modes are typically encountered in regions of the
parameter space with light singlet-like scalars, pseudoscalars, and neutralinos.
$W^{\pm}$ and $Z$ bosons can decay with prompt {or} displaced multi-leptons/taus/jets/photons at the final states.
To detect them, one needs to adopt dedicated experimental searches. Concerning statistics and accuracy, the upcoming colliders (for reviews, see e.g. 
Refs.~\cite{Poeschl:2018etu,Shiltsev:2019rfl}) would proficiently constrain the concerned regions of the parameter space giving rare signals.

\subsubsection{Gravitino LSP}

In Subsec.~\ref{gravitino}, we will discuss that the gravitino LSP is a good candidate for decaying DM which could be detected in gamma-ray experiments.
What we want to point out here is that in the framework of gravitino LSP, each particle candidate for LSP analyzed in previous sections would in fact be the NLSP.
Nevertheless, the analysis of their phenomenology at the LHC is not altered, since they decay
into ordinary particles using the same channels as if they were the LSP.
It was checked in Ref.~\cite{Fidalgo:2011ky} that the effect of the decay of the NLSP into gravitino LSP is negligible, with a decay length of the order of km unless the gravitino mass is very low, less than 10 keV.
Thus the results discussed up to now for 
different sparticles {as LSPs can also be applied 
to the case when they are the NLSPs} with the gravitino as the LSP.

\subsection{Non-Collider Experiments}

Here we discuss briefly several non-collider experiments, where the $\mn$ phenomenological characteristics can be relevant for current data.

\subsubsection{Muon $g-2$}

One of the long standing problems of the SM is the $3.7\,\sigma$ deviation between the measured value of the muon anomalous magnetic dipole moment, $a_\mu = (\text{g}-2)_\mu /2$,
and its theoretical prediction (for a recent review, see Ref.~\cite{Aoyama:2020ynm}). In the framework of the $\mu\nu$SSM, light muon left sneutrino and wino masses can be naturally obtained driven by neutrino physics. This produces an increase of
the dominant chargino-sneutrino loop contribution to muon $g-2$, solving the gap between the theoretical computation and the experimental data. 
The parameter space was analyze in Ref.~\cite{Kpatcha:2019pve}, sampling the $\mu\nu$SSM
with a likelihood data-driven method, paying special attention to reproduce the current experimental data on neutrino and Higgs physics, as well as on flavor observables. 
In addition, the constraints from LHC searches for events with {multi-leptons} + MET were applied on the viable regions found. 
They can probe this scenario through chargino/chargino and chargino/neutralino pair production.

\subsubsection{Neutrino Experiments} 

Neutrino physics is a key {subject} of the $\mn$. As discussed throughout the review,
neutrino masses and mixing angles can easily {be reproduced through}
a generalized EW seesaw, and, in addition, CP phases in the neutrino sector can be accommodated even with real parameters~\cite{Fidalgo:2009dm}.  
On the other hand, given that sterile neutrino searches is an active issue~\cite{Arguelles:2019xgp}, what
we would like to point out here is that the $\mn$ can also contribute to it.
In Subsec.~\ref{thenumber}, we saw that the number of RH neutrinos in the $\mn$ in not fixed.
Less or more than three are possible. Thus, with {an} appropriate election of Yukawa parameters,
some of these RH neutrinos might behave as sterile neutrinos giving rise to an interesting phenomenology. See also the discussion in Sec.~\ref{dmneutrinos} concerning their potential contribution to cosmology.




\section{Cosmology}
\label{sec:cosmo}

Relevant cosmological problems of the SM can be attacked in the context of the $\mn$.
The aim of this section is to review them.
First, we will discuss briefly a solution to the generation of the observed 
baryon-antibaryon asymmetry of the Universe through electroweak baryogenesis.
Then, in the second subsection, we will analyze in some detail the accommodation of the
gravitino and/or axino LSP as eligible DM candidates. We will also discuss briefly a proposal for other potential DM candidates from the neutrino sector.  
Finally, the last subsection will be devoted to discuss the cosmological domain wall problem that might be present in the model under certain circumstances, as well as its possible solutions.


\subsection{Electroweak Baryogenesis}

A popular solution to the problem of the baryon asymmetry of the Universe is to use the mechanism of 
thermal leptogenesis. However,
the $\mn$ does not allow a conventional thermal leptogenesis scenario because of the low-scale seesaw used~\cite{Farzan:2005ez}, since we are working with a seesaw at the EW scale. Nevertheless, it was proved in Ref.~\cite{Chung:2010cd} that the EW phase transition
in the $\mn$ is sufficiently strongly first order such that the created baryons are not washed out, realizing {electroweak baryogenesis}. The main reason for this being the extended Higgs sector of the model which includes in the analysis of the minimum of the scalar potential the contribution of {\bf VEVs from right sneutrinos}.

\subsection{Dark Matter}


\subsubsection{Gravitino}
\label{gravitino}

Embedding the $\mn$ in the framework of supergravity, one can accommodate 
the {\bf gravitino LSP as a decaying DM candidate} with a life-time greater than the age of the Universe.
The key to the above is that gravitino interactions are suppressed not only by the Planck scale, but also by the small RPV parameters controlled by the neutrino Yukawas. Thus, {\bf the connection between neutrino physics and RPV is crucial for the cosmology} of the model.
The detection of gravitino DM through the observation of gamma-ray lines (and a smooth spectral signature) in the {\it Fermi} satellite was analyzed in Refs.~\cite{Choi:2009ng,GomezVargas:2011ph,Albert:2014hwa,GomezVargas:2017,Gomez-Vargas:2019mqk}.
The prospects related to future gamma-ray space missions such as e-ASTROGAM and AMEGO were also discussed recently in Ref.~\cite{Gomez-Vargas:2019mqk}.


\begin{figure}[]
 \begin{center}
       \epsfig{file=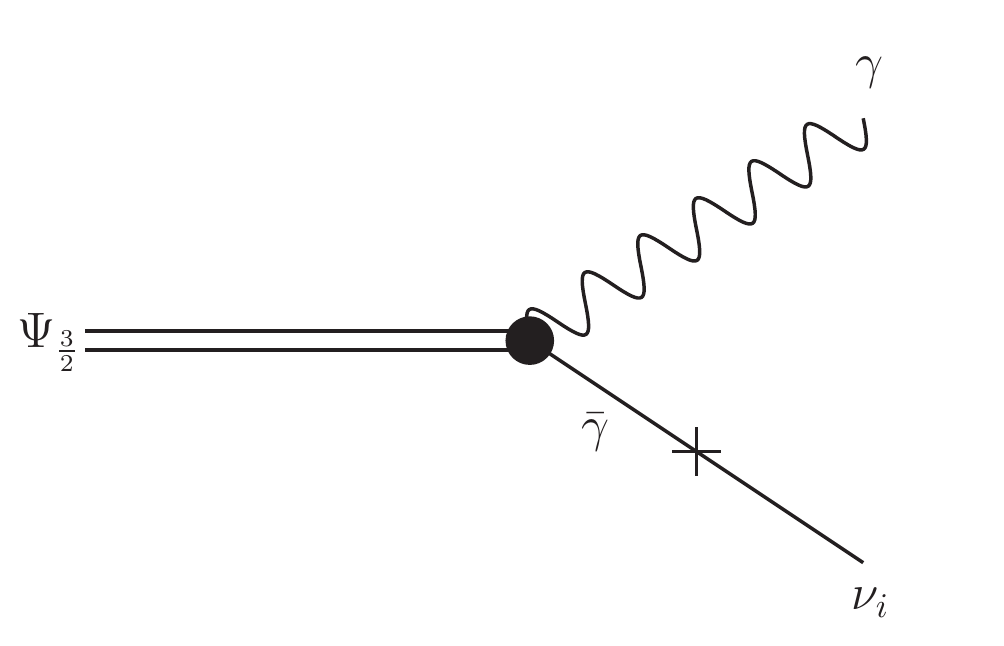,height=4.1cm}
\captions{Tree-level diagram for the two-body decay of a gravitino into a photon and a neutrino, via 
photino-neutrino mixing.}
    \label{fig_gfn}
 \end{center}
\end{figure}

The gravitino is the superpartner of the graviton, and therefore is always present when global SUSY is made local, generating supergravity (for a review, see e.g. Ref.~\cite{Nilles:1983ge}).
The gravitino has in the supergravity Lagrangian an interaction term with photon and photino. In the presence of RPV, photino and LH neutrinos are mixed in the neutral fermion mass matrix, as discussed in Subsec.~\ref{neutralinos}. {Therefore the gravitino LSP is able to decay into photon and neutrino through this interaction term, as shown in Fig.~\ref{fig_gfn}.}

The decay of the gravitino has significant implications because the 
signals are gamma-ray lines with energies half of the gravitino mass 
$m_{3/2}$, that could be detected in gamma-ray satellite experiments.
Gravitino decay width into photon-neutrino through RPV couplings is given by 
\cite{Borgani:1996ag,Takayama:2000uz}:
\bea
\Gamma(\Psi_{3/2}\rightarrow\gamma\nu_i)\simeq\frac{m_{3/2}^3}{32\pi M_{P}^2}|U_{\tilde{\gamma} \nu}|^2,
\label{decay2bodygravitino}
\eea
where $\Gamma(\Psi_{3/2}\rightarrow\gamma\nu_i)$ denotes a sum of the partial decay widths into $\nu_i$ and $\overline{\nu}_i$, 
$M_P\approx 2.43 \times 10^{18}$ GeV is the reduced Planck mass, and
the mixing parameter $U_{\tilde{\gamma} \nu}$ determines the photino content of the neutrino, 
\bea
\left|U_{\tilde{\gamma} \nu}\right|^2= \sum^3_{i=1}\left|N_{i1} \, \cos\theta_W +  N_{i2} \, \sin\theta_W\right|^2.
\label{photino}
\eea
Here $N_{i1} (N_{i2})$ is the bino (wino) component of the $i$-th neutrino, and $\theta_{W}$ is the weak mixing angle.
One can easily estimate the value of $|U_{\widetilde{\gamma}\nu}|$
in the $\mu\nu$SSM~\cite{Choi:2009ng}. Using the mass insertion technique,
from the entries in the neutral fermion mass matrix 
and Fig.~\ref{fig_gfn},
one can deduce that the relevant coupling for the mixing between the photino and the neutrinos is given approximately by $g'v_{iL}$, and as a consequence
\begin{equation}
|U_{\widetilde{\gamma}\nu}|\sim \frac{g'v_{iL}}{M_{1}}.
\label{representa}
\end{equation}
For typical electroweak-scale values for $M_1$, and 
$v_{iL}\lsim 10^{-4}$ GeV as discussed in Subsec.~\ref{sec:higgs potential},
one obtains approximately that 
the photino-neutrino mixing parameter is in the range
\begin{equation}
10^{-8} \lesssim |U_{\widetilde{\gamma}\nu}| \lesssim 10^{-6}.
\label{relaxing}
\end{equation}
This was confirmed in Refs.~\cite{Choi:2009ng,GomezVargas:2017} performing scans in the low-energy parameters of the $\mn$ in order to reproduce the observed neutrino masses and mixing angles.
Relaxing some of the assumptions such as an approximate GUT relation for gaugino masses and/or TeV scales for them, the lower bound can even be smaller:
$10^{-10} \lesssim |U_{\widetilde{\gamma}\nu}| \lesssim 10^{-6}$.
As we can see from Eq.~(\ref{decay2bodygravitino}), the gravitino decay is suppressed both, by the small RPV mixing parameter $|U_{\widetilde{\gamma}\nu}|$, and by the scale of the gravitational interaction, making its lifetime much longer than the age of the Universe $\tau_{3/2}\gg t_{today}\sim 10^{17}$ s, with
\begin{equation}
{\tau}_{3/2}=\Gamma^{-1}(\psi_{3/2}\rightarrow\gamma\nu_i)
\simeq 3.8\times 10^{33}\, {s}
\left(\frac{10^{-8}}{|U_{\widetilde{\gamma}\nu}|}\right)^2
\left(\frac{0.1\, \mathrm{GeV}}{m_{3/2}}\right)^{3}\ .
\label{gravitinolifetime}
\end{equation}

In addition to the two-body decay shown in Fig.~\ref{fig_gfn}, producing an
anisotropic sharp line, there are three-body decays producing a smooth spectral signature.
The gravitino decays into a fermion-antifermion pair and a neutrino via an intermediate photon or $Z$ boson, or into two fermions and a charged lepton, via an intermediate $W^{\pm}$ boson. All these processes 
 were included in the analysis of 
Ref.~\cite{GomezVargas:2017}, where a deep exploration of the low-energy parameter space of the $\mn$ was performed 
in order to compare the gamma-ray fluxes predicted by the model with \Fermi LAT observations.
The result of the analysis implies that to avoid too large fluxes, gravitino masses must be smaller than about 17 GeV and life-times larger
than $4\times 10^{25}$ s.

\subsubsection{Axino}


The axino is the superpartner of the axion, which is a very well motivated particle candidate for solving the strong CP problem of the SM.
The {axino LSP is another decaying DM candidate} in the $\mn$ in a similar way to the gravitino, since it has also an interaction term with
photon and photino.
The interaction is suppressed by the small RPV parameters as in the gravitino case, and by the Peccei-Quinn scale instead of the gravitational one.
This is sufficient to have the axino with a life-time greater than the age of the Universe, but producing a signal with a gamma-ray line with energy half of the axino mass.

In the case of the most popular axion models, 
KSVZ~\cite{Shifman:1979if,Brandenburg:2004du} and 
DFSZ~\cite{Dine:1981rt,Zhitnitsky:1980tq}, this scenario was analyzed in 
Ref.~\cite{Gomez-Vargas:2019vci}. 
The conclusions using \Fermi LAT constraints are that axino masses must be smaller than about 3 GeV. In addition, it was found that a significant region of the parameter space of axino DM lies
in the ballpark of future gamma-ray missions such as the proposed e-ASTROGAM, allowing to explore masses and lifetimes in the ranges $2$ MeV$-3$ GeV 
and $2 \times 10^{26}-8 \times 10^{30}$ s, respectively.

\subsubsection{RH Neutrinos}

\label{dmneutrinos}

We discussed in Subsec.~\ref{thenumber} that the number of RH neutrinos is in general a free parameter in the $\mn$. What we would like to point out here is that 
some of them might behave as sterile neutrinos and be viable 
candidates for warm DM (for a review, see e.g. Ref.~\cite{Boyarsky:2018tvu}). This is similar to the situation of the (non-SUSY) neutrino minimal standard model ($\nu$MSM), where sterile neutrinos play the role of DM~\cite{Asaka:2005pn}. 
To achieve this in the case of the $\mn$, we need some of the RH neutrinos to have small couplings, $Y^\nu \sim 10^{-13}$ and $\kappa\sim 10^{-8}$, in such a way that they obtain keV masses, and lifetimes long enough to be candidates for DM~\cite{knees}.
Let us finally point out that the other possible compositions of the $\mn$ neutralino are not convenient for DM, since one would need extremely small EW gaugino masses of the order of the keV, or alternatively Higgsino masses also of that order.
This is not possible for Higgsinos and winos because the LEP bound on the lightest chargino mass imposes $\mu, M_2\gsim$ 94 GeV.
It is not natural for binos whose soft mass $M_1$ is expected to be of the order of the TeV.
Besides, using the GUT relation one obtains a lower bound on $M_1$, and therefore on the lightest neutralino mass of 46 GeV~\cite{Tanabashi:2018oca}.



\subsubsection{Multicomponent DM}

In Refs.~\cite{Gomez-Vargas:2019mqk,Gomez-Vargas:2019vci},
the interesting possibility of a {multicomponent DM scenario} made of gravitino and axino was
analyzed.
If the axino is the LSP, interestingly a gravitino NLSP can live enough as to contribute to the relic density. It was found that the axino or the gravitino can produce a signal detectable by future MeV-GeV gamma-ray telescopes such as e-ASTROGAM. In addition, there is a parameter region
where a well-tempered mixture of both particles is obtained, with a double-line signal arising as a smoking gun.
Similar qualitative conclusions are obtained in the opposite case with gravitino LSP and axino NLSP. 
Of course, other DM candidates could also contribute to the total amount of DM, such as the sterile neutrinos discussed above, the well-known axion, etc.

\vspace{0.25cm}

\noindent
Summarizing the above discussions, the $\mn$ is an appealing scenario for solving the DM problem, with different interesting potential candidates.
If several of them contribute to DM at the same time, {this might relax potential tensions
between the standard $\Lambda$CDM model and cosmological observations \cite{Berezhiani:2015,Chudaykin:2016,Poulin:2016,Chudaykin:2017,Bringmann:2018}.}

\subsection{The Domain Wall Problem}
\label{domain}

As discussed in Subsec.~\ref{proton}, the superpotential of the $\mn$ in Eq.~(\ref{superpotential}) has a $Z_3$ symmetry.
This discrete symmetry can induce a cosmological domain wall 
problem~\cite{Holdom:1983vk,Ellis:1986mq,Rai:1992xw,Abel:1995wk,Chung:2010cd}, unless inflation at the weak scale is invoked. This problem, if present, {\bf can be solved with the presence 
of non-renormalizable operators} in $W$.
These operators break explicitly the $Z_3$ symmetry lifting the degeneracy of the three original vacua, and they can be chosen small enough as not to alter the low-energy 
phenomenology~\cite{Holdom:1983vk,Ellis:1986mq,Rai:1992xw,Chung:2010cd}. 
It is true that 
in the context of supergravity they can reintroduce in $W$ the linear and bilinear terms forbidden by the $Z_3$ symmetry~\cite{Abel:1995wk}, and generate quadratic tadpole divergences~\cite{Ellwanger:1983mg,Bagger:1993ji,Jain:1994tk,Bagger:1995ay,Abel:1996cr}, nevertheless
these problems can be eliminated in models which possess a
$R$-symmetry in the non-renormalizable superpotential~\cite{Abel:1996cr,Panagiotakopoulos:1998yw}.

As already mentioned in Subsec.~\ref{proton},
another strategy to solve this potential problem is through
{\bf the addition of an extra $U(1)'$ gauge symmetry} in the $\mn$~\cite{Fidalgo:2011tm}.
As it is well known, the domain wall problem disappears
once the discrete symmetry is embedded in the gauge 
symmetry~\cite{Lazarides:1982tw,Kibble:1982dd,Barr:1982bb}.

\section{Extensions}
\label{sec:extensions}

There has been studies of extensions of the $\mn$ in the literature, interesting both from the phenomenological and cosmological viewpoints. In this section we will briefly describe their main characteristics and why they can be useful to solve several potential problems.

\subsection{Trilinear Lepton-Number Violating Couplings}
\label{trilinearl}

The presence of the {conventional} lepton-number violating couplings $\lambda_{ijk} \hat L_i \hat L_j \hat e^c_k$ and $\lambda'_{ijk}\hat L_i \hat Q_{j} \hat d^{c}_{k}$ is expected naturally in the
$\mn$.
This issue was discussed in detail in Refs.~\cite{Lopez-Fogliani:2017qzj,Ghosh:2017yeh}.
The reason is the following. Given that the superfields $\hat H_d$ and $\hat L_i$ have the same gauge quantum numbers, once the usual Yukawa couplings for $d$-type quarks $Y^d_{ij}$ and charged leptons $Y^e_{ij}$ are introduced in $W$, exchanging $\hat H_d\rightarrow \hat L_i$ the presence of the lepton-number violating couplings is natural. This is similar to the presence of the 
effective $\mu$-term in $W$ determined by $\lambda_i$, once Yukawa couplings for neutrinos $Y^{\nu}_{ij}$ are added, exchanging in this case $\hat L_i\rightarrow \hat H_d$.
Thus, we can write an extension of the superpotential of Eq.~(\ref{superpotential}) as
\bea
W = &&
\epsilon_{ab} \left(
Y^e_{ij} \, \hat H_d^a\, \hat L^b_i \, \hat e_j^c +
Y^d_{ij} \, \hat H_d^a\, \hat Q^{b}_{i} \, \hat d_{j}^{c} 
+
Y^u_{ij} \, \hat H_u^b\, \hat Q^{a}_{i} \, \hat u_{j}^{c}
\right)
\nonumber\\
&+&   
\epsilon_{ab} \left(\lambda_{ijk} \hat L_i^a \hat L_j^b \hat e^c_k 
+ 
 \lambda'_{ijk} \hat L_i^a \hat Q_{j}^{b} \hat d^{c}_{k} \right)
\nonumber\\
&+&   
\epsilon_{ab} \left(
Y^{\nu}_{ij} \, \hat H_u^b\, \hat L^a_i \, \hat \nu^c_j 
-
\lambda_{i} \,  \hat H_u^b \hat H_d^a \,  \hat \nu^c_i
\right)
+
\frac{1}{3}
\kappa{_{ijk}} 
\hat \nu^c_i\hat \nu^c_j\hat \nu^c_k
\ ,
\label{superpotentiallb}
\eea
where the new terms are written in the second line.

As discussed in Subsec.~\ref{comparison}, these terms are the conventional TRPV couplings.
Their quadratic coupling constant products 
$\lambda_{ijk} \lambda_{lmn}$, $\lambda_{ijk}\lambda'_{lmn}$ 
and $\lambda'_{ijk}\lambda'_{lmn}$ 
are strongly constrained 
(see e.g. Refs.~\cite{Chemtob:2004xr,Barbier:2004ez,Dreiner:2012mx} for reviews).
The approach adopted in previous sections has been to neglect the values of the individual couplings in case they are present, in such a way that the superpotential of Eq.~(\ref{superpotentiallb}) can be approximated by the one of Eq.~(\ref{superpotential}). 
It is true that even if 
$\lambda_{ijk}$ and $\lambda'_{ijk}$ are not present at tree level, they will appear through loop 
processes~\cite{Escudero:2008jg,Ghosh:2017yeh}, however their contributions are
smaller than order $10^{-9}$. 
Thus, although
all existing bounds on quadratic coupling constant products 
are satisfied, these contributions are anyway negligible for studying physical processes.

However, if not all of these tree-level couplings are negligible and some of them are comparable to the distinctive couplings of the $\mn$, the physical processes discussed in Sec.~\ref{sec:searches} can be modified and new signals could be present. 
This deserves further studies.



\subsection{Extra $U(1)'$ Gauge Symmetry}
\label{gauge}

We have argued in Subsec.~\ref{proton} that string theory could be the source of the effective $Z_3$ discrete symmetry of the superpotential in Eq.~(\ref{superpotential}). 
This symmetry is useful not only to forbid the presence of the dangerous linear (tadpole) terms $t_{i} \hat\nu_{i}^c$ in the superpotential, but also the presence of the bilinear (mass) terms
$\mu\hat H_u \hat H_d$, $\mu_i \hat H_u \hat L_i$ and 
${\mathcal M}_{ij} \hat\nu_{i}^c\hat\nu_{j}^c$. The latter mass terms
would reintroduce  
the $\mu$-problem and additional naturalness problems.
Alternatively, another interesting strategy to forbid all these dangerous operators, including the baryon-number violating couplings $\lambda''_{ijk}\hat d^c_{i} \hat d^c_{j} \hat u^c_{k}$, as well as to avoid the cosmological domain wall problem generated by the $Z_3$ symmetry, it to add an extra $U(1)'$ gauge group.

This mechanism was first adopted in Ref.~\cite{Fidalgo:2011tm} in the context of the $\mn$, 
extending therefore the SM gauge group to
$SU(3)\times SU(2)\times U(1)_Y\times U(1)'$
(see Ref.~\cite{Langacker:2008yv} for a review in other models).
The fields are in general charged under this new symmetry making the dangerous terms not gauge invariant, and therefore forbidden.
Needless to say, a new $Z'$ gauge boson is present in the spectrum giving rise to a rich phenomenology.
Also, extra matter in the form of color triplets, electroweak doublets and SM singlets is typically present in the spectrum, because of the anomaly cancellation conditions. For alternative models without exotic matter, thanks to
allow for non-universal $U(1)'$ charges of the SM fields, see
Ref.~\cite{Lozano:2018esg}. 

It is worth noticing that the presence of the extra $U(1)'$ forbids the usual
effective Majorana mass terms $\frac{1}{3}
\kappa{_{ijk}} 
\hat \nu^c_i\hat \nu^c_j\hat \nu^c_k$ in the superpotential of the original $\mn$.
Thus, in the case of assuming three families of RH neutrinos they can only acquire large masses
through the mixing with the extra gaugino and the Higgsinos.
Then, only two RH neutrinos can have EW-scale masses, and the third one combines with the LH neutrinos to form a nearly massless Dirac particle. 
As a consequence the EW seesaw only works for two linear combinations of LH neutrinos.
At tree level there are four light Majorana states. To account for neutrino data some of the
entries of the Yukawa matrix must be very small $Y^\nu \lsim 10^{-11}$.
At the end of the day one obtains two heavy RH neutrinos
of the order of TeV and four light (three active and one sterile) neutrinos~\cite{Fidalgo:2011tm}.

Although in the models built in Ref.~\cite{Fidalgo:2011tm}, in addition to three RH neutrinos three families of extra matter were also imposed in similarity with the characteristics of the SM spectrum, this is not a necessary condition.
The fact of not imposing three families of SM singlets as well as of extra matter gives rise to a different and interesting phenomenology/cosmology~\cite{lopez:2020xx}. In addition to exotic quarks, a number of RH neutrinos different from three can arise compatible with anomaly cancellations,
{as well as new singlets under the SM gauge group.
Some of the latter superfields ($\hat S$) can generate Majorana masses for RH neutrinos through couplings of the type $\hat S \hat \nu^c \hat \nu^c$, and others ($\hat\chi$) will be DM 
candidates through couplings $\hat \nu^c \hat \chi \hat \chi$.
After EWSB the $\hat\chi$ fields acquire EW-scale masses of the order of $v_{R}$.
Because of the $Z_2$ symmetry of the coupling, the lightest of the scalar and fermionic components can behave as {\bf stable WIMP DM}~\cite{lopez:2020xx}, even though we are working in the context of a model with RPV. This is an intriguing possibility which deserves further research. 
}

\subsection{Reinterpretation of the Higgs Field}


A reinterpretation of the Higgs field in SUSY was proposed in Ref.~\cite{Lopez-Fogliani:2017qzj}.
The structure of the $\mn$, where Higgs and lepton superfields are mixed, was the motivation for this proposal.
In particular, it was advocated to interpret the two Higgs 
doublets as a fourth family of lepton superfields. 
This 
motivates the possibility of the existence of a fourth family of {\bf vector-like quark doublets}, whose new signals were studied in detail in
Ref.~\cite{Aguilar-Saavedra:2017giu}.




\section{Conclusions and Outlook}
\label{conclusions}

We have reviewed throughout this work the $\mn$~\cite{LopezFogliani:2005yw}. 
This supersymmetric model was proposed in 2005 as an alternative to the MSSM and other supersymmetric models existing in the literature. 
We have tried to explain that several phenomenological and cosmological problems of the standard model might have explanations within {the $\mn$} framework: {from neutrino physics to dark matter and
baryogenesis}, through the detection of new physics.
The main characteristic of the model, and hence its simplicity, is to include the right-handed neutrinos in the spectrum. After all, now we know that neutrinos are massive. This small change, allows to explain 
the $\mu$- and $\nu$-problems without relying in new fields for this task, or new scales in addition to the {electroweak} one.

It is true that the presence of right-handed neutrinos implies automatically that $R$-parity is violated.
But this that seemed like a problem years ago, now seems more like a virtue, by avoiding the stringent bounds on supersymmetric partners from experimental data.
In fact, because of these results, the analysis of displaced signals at the LHC begins to be fashionable (for a recent review, see Ref.~\cite{Alimena:2019zri}).
With this work, we have tried to convince the reader that it is not yet time to throw in the towel and abandon supersymmetry. In a model like the $\mn$, the current analyses only constrain its parameter space very mildly. 
However, its predictions can be very exciting, since the smallness of neutrino masses is directly connected with the low decay width of the LSP. 
Thus, neutrino physics and supersymmetry are inseparably related.

{For all of the above, we consider that analyzing the (collider) phenomenology of the model is a crucial task for the future in the framework of the $\mn$. Analyses of the electroweak sector have already been carried out, such as detail studies of left sleptons with the sneutrino as the LSP~\cite{Ghosh:2017yeh,Lara:2018rwv,Kpatcha:2019gmq}.
After recasting the specific displaced-vertex ATLAS search of Ref.~\cite{Aad:2015rba}, the result~\cite{Kpatcha:2019gmq} is that the only parameter-independent {bound} comes from Higgs data, establishing that left slepton masses must be heavier than half of the mass of the standard-model-like Higgs. Nevertheless, significant regions with a tau left sneutrino LSP could be probed at the LHC run 3 as summarized in Fig.~\ref{Cut-MSvL3-M-Av3.png}. 
On the other hand, most of the parameter space of the neutralino as LSP, and specially the right-handed neutrino-like LSP which is genuine of the $\mn$, has not been compared with LHC results yet. 
After recasting the specific prompt ATLAS searches of Refs.~\cite{Aaboud:2018sua,Aad:2019vvi}, only a small
region with bino LSP (and sneutrino NLSP) of mass $110-150$ ($110-160$) GeV was 
excluded~\cite{Lara:2018zvf} (see Fig.~\ref{fig:exclusions}).
Besides, the color sector of the $\mn$ has not been analyzed at all.
For example, a stop LSP decays into top plus neutrino and bottom plus lepton, and these
decays can be prompt or displaced depending on the region of the parameter space analyzed,
giving rise to potentially interesting signals that could be constrained with LHC data.
The Higgs sector seems to be also promising, with the possibility of light states below the standard-model-like Higgs. In other words, there is still a lot of work ahead.}

{However, carrying it out is not easy, because one of the problems with existing searches is that they are designed for a generic purpose and therefore not optimized for the type of spectrum and couplings present in the $\mn$. Recasting ATLAS and CMS searches is crucial, and dedicated analyses like the few that have already been carried out are mandatory.}
We believe that the strategy that we have already started to employ is adequate.
In particular, we are analyzing signals of new physics at the LHC predicted by the model, performing scans of the parameter space with a powerful likelihood data-driven method. Imposing compatibility with current experimental data on neutrino and Higgs physics as well as with flavor observables, we expect to prove the $\mn$ in the near future.

{Less peremptory but not less important, is to continue analyzing the cosmology of the model, paying special attention to the detection of the possible candidates for dark matter that the model contains, as well as to the reanalysis of the electroweak 
{baryogenesis} mechanism to generate the baryonic asymmetry of the Universe, in light of the experimental results 
{about} Higgs physics. Finally, incorporating inflation into the model is an interesting task, and especially if it is possible to achieve it at the electroweak scale so that it is still the only scale of our theory.}
{Nor should we forget non-collider experiments. 
The continue improvements in neutrino physics experiments, as for instance the measurements of CP violating phases or the search for sterile neutrinos, are crucial nowadays and the $\mn$ has much to say about them given the relevant role played by the neutrinos in the model.
Also, we look forward to the new results from muon $g-2$ experiments.
In the event that the discrepancy between theory and experiment continues, the new physics of the
$\mn$ might contribute to solving the puzzle.}

%

\section*{Acknowledgments}
{Some of the results discussed here were obtained in cooperation with the following collaborators to whom we are very grateful: J.A. Aguilar-Saavedra, T. Biekotter, K.-Y. Choi, N. Escudero, J. Fidalgo, P. Ghosh, G.A. G\'omez-Vargas, S. Heinemeyer, E. Kpatcha, I. Lara, V.A. Mitsou, N. Nagata, H. Otono, A.D. P\'erez and R. Ruiz de Austri.}
The work of DL was supported by the Argentinian CONICET, and also acknowledges the support through PIP 11220170100154CO. 
The work of CM was supported in part by the Spanish Agencia Estatal de Investigaci\'on 
through the grants 
PGC2018-095161-B-I00 and IFT Centro de Excelencia Severo Ochoa SEV-2016-0597.
We also acknowledge the support of the Spanish Red Consolider MultiDark FPA2017-90566-REDC. 

\noindent 
{\bf Author contribution statement}:
Both authors, DL and CM, contributed equally to write this review.


\bibliographystyle{utphys}
\bibliography{munussmbib-completo}

\end{document}